\documentclass[a4paper,traditabstract,longauth]{aa} % for the abstract without uration 
\usepackage{graphicx}
\usepackage{amsmath}
\usepackage[breaklinks, colorlinks, citecolor=blue]{hyperref}
\usepackage{fixltx2e}
\usepackage{natbib}
\usepackage{txfonts}
\usepackage{ifthen}

\usepackage{hyperref}
\usepackage{bookmark}

\usepackage[switch]{lineno}
\newcommand{\XMM}{XMM}
\newcommand{\CHANDRA}{\textit{Chandra}}
\newcommand{\ROSAT}{ROSAT}

\usepackage{color}

\def\setsymbol#1#2{\expandafter\def\csname #1\endcsname{#2}}
\def\getsymbol#1{\csname #1\endcsname}

\def\Planck{\textit{Planck}}
\newbox\tablebox    \newdimen\tablewidth
\def\leaderfil{\leaders\hbox to 5pt{\hss.\hss}\hfil}
\def\endPlancktablewide{\tablewidth=\textwidth 
    $$\hss\copy\tablebox\hss$$
    \vskip-\lastskip\vskip -2pt}
\def\tablenote#1 #2\par{\begingroup \parindent=0.8em
    \abovedisplayshortskip=0pt\belowdisplayshortskip=0pt
    \noindent
    $$\hss\vbox{\hsize\tablewidth \hangindent=\parindent \hangafter=1 \noindent
    \hbox to \parindent{$^#1$\hss}\strut#2\strut\par}\hss$$
    \endgroup}
\def\doubleline{\vskip 3pt\hrule \vskip 1.5pt \hrule \vskip 5pt}

\def\L2{\ifmmode L_2\else $L_2$\fi}

\def\DeltaT{\ifmmode \Delta T\else $\Delta T$\fi}
\def\deltat{\ifmmode \Delta t\else $\Delta t$\fi}
\def\fknee{\ifmmode f_{\rm knee}\else $f_{\rm knee}$\fi}
\def\Fmax{\ifmmode F_{\rm max}\else $F_{\rm max}$\fi}
\def\solar{\ifmmode{\rm M}_{\mathord\odot}\else${\rm M}_{\mathord\odot}$\fi}
\def\Msolar{\ifmmode{\rm M}_{\mathord\odot}\else${\rm M}_{\mathord\odot}$\fi}
\def\Lsolar{\ifmmode{\rm L}_{\mathord\odot}\else${\rm L}_{\mathord\odot}$\fi}
\def\inv{\ifmmode^{-1}\else$^{-1}$\fi}
\def\mo{\ifmmode^{-1}\else$^{-1}$\fi}
\def\sup#1{\ifmmode ^{\rm #1}\else $^{\rm #1}$\fi}
\def\expo#1{\ifmmode \times 10^{#1}\else $\times 10^{#1}$\fi}
\def\,{\thinspace}
\def\lsim{\mathrel{\raise .4ex\hbox{\rlap{$<$}\lower 1.2ex\hbox{$\sim$}}}}
\def\gsim{\mathrel{\raise .4ex\hbox{\rlap{$>$}\lower 1.2ex\hbox{$\sim$}}}}

\def\simprop{\mathrel{\raise .4ex\hbox{\rlap{$\propto$}\lower 1.2ex\hbox{$\sim$}}}}
\def\deg{\ifmmode^\circ\else$^\circ$\fi}
\def\pdeg{\ifmmode $\setbox0=\hbox{$^{\circ}$}\rlap{\hskip.11\wd0 .}$^{\circ}
          \else \setbox0=\hbox{$^{\circ}$}\rlap{\hskip.11\wd0 .}$^{\circ}$\fi}
\def\arcs{\ifmmode {^{\scriptstyle\prime\prime}}
          \else $^{\scriptstyle\prime\prime}$\fi}
\def\arcm{\ifmmode {^{\scriptstyle\prime}}
          \else $^{\scriptstyle\prime}$\fi}
\newdimen\sa  \newdimen\sb
\def\parcs{\sa=.07em \sb=.03em
     \ifmmode \hbox{\rlap{.}}^{\scriptstyle\prime\kern -\sb\prime}\hbox{\kern -\sa}
     \else \rlap{.}$^{\scriptstyle\prime\kern -\sb\prime}$\kern -\sa\fi}
\def\parcm{\sa=.08em \sb=.03em
     \ifmmode \hbox{\rlap{.}\kern\sa}^{\scriptstyle\prime}\hbox{\kern-\sb}
     \else \rlap{.}\kern\sa$^{\scriptstyle\prime}$\kern-\sb\fi}
\def\ra[#1 #2 #3.#4]{#1\sup{h}#2\sup{m}#3\sup{s}\llap.#4}
\def\dec[#1 #2 #3.#4]{#1\deg#2\arcm#3\arcs\llap.#4}
\def\deco[#1 #2 #3]{#1\deg#2\arcm#3\arcs}
\def\rra[#1 #2]{#1\sup{h}#2\sup{m}}

\def\dots{\relax\ifmmode \ldots\else $\ldots$\fi}
\def\WHzsr{\ifmmode $W\,Hz\mo\,sr\mo$\else W\,Hz\mo\,sr\mo\fi}
\def\mHz{\ifmmode $\,mHz$\else \,mHz\fi}
\def\GHz{\ifmmode $\,GHz$\else \,GHz\fi}
\def\mKs{\ifmmode $\,mK\,s$^{1/2}\else \,mK\,s$^{1/2}$\fi}
\def\muKs{\ifmmode \,\mu$K\,s$^{1/2}\else \,$\mu$K\,s$^{1/2}$\fi}
\def\muKRJs{\ifmmode \,\mu$K$_{\rm RJ}$\,s$^{1/2}\else \,$\mu$K$_{\rm RJ}$\,s$^{1/2}$\fi}
\def\muKHz{\ifmmode \,\mu$K\,Hz$^{-1/2}\else \,$\mu$K\,Hz$^{-1/2}$\fi}
\def\MJysr{\ifmmode \,$MJy\,sr\mo$\else \,MJy\,sr\mo\fi}
\def\MJysrmK{\ifmmode \,$MJy\,sr\mo$\,mK$_{\rm CMB}\mo\else \,MJy\,sr\mo\,mK$_{\rm CMB}\mo$\fi}
\def\microns{\ifmmode \,\mu$m$\else \,$\mu$m\fi}

\def\muK{\ifmmode \,\mu$K$\else \,$\mu$\hbox{K}\fi}
\def\microK{\ifmmode \,\mu$K$\else \,$\mu$\hbox{K}\fi}
\def\muW{\ifmmode \,\mu$W$\else \,$\mu$\hbox{W}\fi}
\def\kms{\ifmmode $\,km\,s$^{-1}\else \,km\,s$^{-1}$\fi}
\def\kmsMpc{\ifmmode $\,\kms\,Mpc\mo$\else \,\kms\,Mpc\mo\fi}
\providecommand{\sorthelp}[1]{}

\begin{document}

   \title{{\Planck} intermediate results. XL. 
          The Sunyaev-Zeldovich signal from the Virgo cluster}
   \titlerunning{SZ in Virgo as seen by \Planck}
%   \author{Planck collaboration}
%   \authorrunning{Planck collaboration}

%   \date{February 13th 2015}

 \abstract{}{}{}{}{} 
% 5 {} token are mandatory
%This author list corresponds to \title{Author list for PIP\_15\_Proj\_5\_1\_Diego}
%Prepared by M. Lopez-Caniego (Marcos.Lopez.Caniego@sciops.esa.int), ESAC/ESA
%This version is from Thu Nov 12 16:55:48 2015 CET
%\subtitle{There are 204 co-authors in this list}
\author{\small
Planck Collaboration: P.~A.~R.~Ade\inst{86}
\and
N.~Aghanim\inst{59}
\and
M.~Arnaud\inst{73}
\and
M.~Ashdown\inst{69, 5}
\and
J.~Aumont\inst{59}
\and
C.~Baccigalupi\inst{85}
\and
A.~J.~Banday\inst{95, 10}
\and
R.~B.~Barreiro\inst{64}
\and
N.~Bartolo\inst{28, 65}
\and
E.~Battaner\inst{97, 98}
% ...
}
\institute{\small
APC, AstroParticule et Cosmologie, Universit\'{e} Paris Diderot, CNRS/IN2P3, CEA/lrfu, Observatoire de Paris, Sorbonne Paris Cit\'{e}, 10, rue Alice Domon et L\'{e}onie Duquet, 75205 Paris Cedex 13, France\goodbreak
\and
African Institute for Mathematical Sciences, 6-8 Melrose Road, Muizenberg, Cape Town, South Africa\goodbreak
\and
Agenzia Spaziale Italiana Science Data Center, Via del Politecnico snc, 00133, Roma, Italy\goodbreak
\and
Aix Marseille Universit\'{e}, CNRS, LAM (Laboratoire d'Astrophysique de Marseille) UMR 7326, 13388, Marseille, France\goodbreak
\and
Astrophysics Group, Cavendish Laboratory, University of Cambridge, J J Thomson Avenue, Cambridge CB3 0HE, U.K.\goodbreak
\and
Astrophysics \& Cosmology Research Unit, School of Mathematics, Statistics \& Computer Science, University of KwaZulu-Natal, Westville Campus, Private Bag X54001, Durban 4000, South Africa\goodbreak
\and
Atacama Large Millimeter/submillimeter Array, ALMA Santiago Central Offices, Alonso de Cordova 3107, Vitacura, Casilla 763 0355, Santiago, Chile\goodbreak
\and
CGEE, SCS Qd 9, Lote C, Torre C, 4$^{\circ}$ andar, Ed. Parque Cidade Corporate, CEP 70308-200, Bras\'{i}lia, DF,Ê Brazil\goodbreak
\and
CITA, University of Toronto, 60 St. George St., Toronto, ON M5S 3H8, Canada\goodbreak
\and
CNRS, IRAP, 9 Av. colonel Roche, BP 44346, F-31028 Toulouse cedex 4, France\goodbreak
\and
California Institute of Technology, Pasadena, California, U.S.A.\goodbreak
\and
Centro de Estudios de F\'{i}sica del Cosmos de Arag\'{o}n (CEFCA), Plaza San Juan, 1, planta 2, E-44001, Teruel, Spain\goodbreak
\and
Computational Cosmology Center, Lawrence Berkeley National Laboratory, Berkeley, California, U.S.A.\goodbreak
\and
DSM/Irfu/SPP, CEA-Saclay, F-91191 Gif-sur-Yvette Cedex, France\goodbreak
\and
DTU Space, National Space Institute, Technical University of Denmark, Elektrovej 327, DK-2800 Kgs. Lyngby, Denmark\goodbreak
\and
D\'{e}partement de Physique Th\'{e}orique, Universit\'{e} de Gen\`{e}ve, 24, Quai E. Ansermet,1211 Gen\`{e}ve 4, Switzerland\goodbreak
\and
Departamento de Astrof\'{i}sica, Universidad de La Laguna (ULL), E-38206 La Laguna, Tenerife, Spain\goodbreak
\and
Departamento de F\'{\i}sica, Universidad de Oviedo, Avda. Calvo Sotelo s/n, Oviedo, Spain\goodbreak
\and
Department of Astrophysics/IMAPP, Radboud University Nijmegen, P.O. Box 9010, 6500 GL Nijmegen, The Netherlands\goodbreak
\and
Department of Physics \& Astronomy, University of British Columbia, 6224 Agricultural Road, Vancouver, British Columbia, Canada\goodbreak
\and
Department of Physics and Astronomy, Dana and David Dornsife College of Letter, Arts and Sciences, University of Southern California, Los Angeles, CA 90089, U.S.A.\goodbreak
\and
Department of Physics and Astronomy, University College London, London WC1E 6BT, U.K.\goodbreak
\and
Department of Physics, Florida State University, Keen Physics Building, 77 Chieftan Way, Tallahassee, Florida, U.S.A.\goodbreak
\and
Department of Physics, Gustaf H\"{a}llstr\"{o}min katu 2a, University of Helsinki, Helsinki, Finland\goodbreak
\and
Department of Physics, Princeton University, Princeton, New Jersey, U.S.A.\goodbreak
\and
Department of Physics, University of California, Santa Barbara, California, U.S.A.\goodbreak
\and
Department of Physics, University of Illinois at Urbana-Champaign, 1110 West Green Street, Urbana, Illinois, U.S.A.\goodbreak
\and
Dipartimento di Fisica e Astronomia G. Galilei, Universit\`{a} degli Studi di Padova, via Marzolo 8, 35131 Padova, Italy\goodbreak
\and
Dipartimento di Fisica e Scienze della Terra, Universit\`{a} di Ferrara, Via Saragat 1, 44122 Ferrara, Italy\goodbreak
\and
Dipartimento di Fisica, Universit\`{a} La Sapienza, P. le A. Moro 2, Roma, Italy\goodbreak
\and
Dipartimento di Fisica, Universit\`{a} degli Studi di Milano, Via Celoria, 16, Milano, Italy\goodbreak
\and
Dipartimento di Fisica, Universit\`{a} degli Studi di Trieste, via A. Valerio 2, Trieste, Italy\goodbreak
\and
Dipartimento di Fisica, Universit\`{a} di Roma Tor Vergata, Via della Ricerca Scientifica, 1, Roma, Italy\goodbreak
\and
Dipartimento di Matematica, Universit\`{a} di Roma Tor Vergata, Via della Ricerca Scientifica, 1, Roma, Italy\goodbreak
\and
Discovery Center, Niels Bohr Institute, Blegdamsvej 17, Copenhagen, Denmark\goodbreak
\and
Discovery Center, Niels Bohr Institute, Copenhagen University, Blegdamsvej 17, Copenhagen, Denmark\goodbreak
\and
European Southern Observatory, ESO Vitacura, Alonso de Cordova 3107, Vitacura, Casilla 19001, Santiago, Chile\goodbreak
\and
European Space Agency, ESAC, Planck Science Office, Camino bajo del Castillo, s/n, Urbanizaci\'{o}n Villafranca del Castillo, Villanueva de la Ca\~{n}ada, Madrid, Spain\goodbreak
\and
European Space Agency, ESTEC, Keplerlaan 1, 2201 AZ Noordwijk, The Netherlands\goodbreak
\and
Facolt\`{a} di Ingegneria, Universit\`{a} degli Studi e-Campus, Via Isimbardi 10, Novedrate (CO), 22060, Italy\goodbreak
\and
Gran Sasso Science Institute, INFN, viale F. Crispi 7, 67100 L'Aquila, Italy\goodbreak
\and
HGSFP and University of Heidelberg, Theoretical Physics Department, Philosophenweg 16, 69120, Heidelberg, Germany\goodbreak
\and
Helsinki Institute of Physics, Gustaf H\"{a}llstr\"{o}min katu 2, University of Helsinki, Helsinki, Finland\goodbreak
\and
INAF - Osservatorio Astrofisico di Catania, Via S. Sofia 78, Catania, Italy\goodbreak
\and
INAF - Osservatorio Astronomico di Padova, Vicolo dell'Osservatorio 5, Padova, Italy\goodbreak
\and
INAF - Osservatorio Astronomico di Roma, via di Frascati 33, Monte Porzio Catone, Italy\goodbreak
\and
INAF - Osservatorio Astronomico di Trieste, Via G.B. Tiepolo 11, Trieste, Italy\goodbreak
\and
INAF/IASF Bologna, Via Gobetti 101, Bologna, Italy\goodbreak
\and
INAF/IASF Milano, Via E. Bassini 15, Milano, Italy\goodbreak
\and
INFN, Sezione di Bologna, Via Irnerio 46, I-40126, Bologna, Italy\goodbreak
\and
INFN, Sezione di Roma 1, Universit\`{a} di Roma Sapienza, Piazzale Aldo Moro 2, 00185, Roma, Italy\goodbreak
\and
INFN, Sezione di Roma 2, Universit\`{a} di Roma Tor Vergata, Via della Ricerca Scientifica, 1, Roma, Italy\goodbreak
\and
INFN/National Institute for Nuclear Physics, Via Valerio 2, I-34127 Trieste, Italy\goodbreak
\and
IPAG: Institut de Plan\'{e}tologie et d'Astrophysique de Grenoble, Universit\'{e} Grenoble Alpes, IPAG, F-38000 Grenoble, France, CNRS, IPAG, F-38000 Grenoble, France\goodbreak
\and
IUCAA, Post Bag 4, Ganeshkhind, Pune University Campus, Pune 411 007, India\goodbreak
\and
Imperial College London, Astrophysics group, Blackett Laboratory, Prince Consort Road, London, SW7 2AZ, U.K.\goodbreak
\and
Infrared Processing and Analysis Center, California Institute of Technology, Pasadena, CA 91125, U.S.A.\goodbreak
\and
Institut Universitaire de France, 103, bd Saint-Michel, 75005, Paris, France\goodbreak
\and
Institut d'Astrophysique Spatiale, CNRS (UMR8617) Universit\'{e} Paris-Sud 11, B\^{a}timent 121, Orsay, France\goodbreak
\and
Institut d'Astrophysique de Paris, CNRS (UMR7095), 98 bis Boulevard Arago, F-75014, Paris, France\goodbreak
\and
Institute of Astronomy, University of Cambridge, Madingley Road, Cambridge CB3 0HA, U.K.\goodbreak
\and
Institute of Theoretical Astrophysics, University of Oslo, Blindern, Oslo, Norway\goodbreak
\and
Instituto de Astrof\'{\i}sica de Canarias, C/V\'{\i}a L\'{a}ctea s/n, La Laguna, Tenerife, Spain\goodbreak
\and
Instituto de F\'{\i}sica de Cantabria (CSIC-Universidad de Cantabria), Avda. de los Castros s/n, Santander, Spain\goodbreak
\and
Istituto Nazionale di Fisica Nucleare, Sezione di Padova, via Marzolo 8, I-35131 Padova, Italy\goodbreak
\and
Jet Propulsion Laboratory, California Institute of Technology, 4800 Oak Grove Drive, Pasadena, California, U.S.A.\goodbreak
\and
Jodrell Bank Centre for Astrophysics, Alan Turing Building, School of Physics and Astronomy, The University of Manchester, Oxford Road, Manchester, M13 9PL, U.K.\goodbreak
\and
Kavli Institute for Cosmological Physics, University of Chicago, Chicago, IL 60637, USA\goodbreak
\and
Kavli Institute for Cosmology Cambridge, Madingley Road, Cambridge, CB3 0HA, U.K.\goodbreak
\and
Kazan Federal University, 18 Kremlyovskaya St., Kazan, 420008, Russia\goodbreak
\and
LAL, Universit\'{e} Paris-Sud, CNRS/IN2P3, Orsay, France\goodbreak
\and
LERMA, CNRS, Observatoire de Paris, 61 Avenue de l'Observatoire, Paris, France\goodbreak
\and
Laboratoire AIM, IRFU/Service d'Astrophysique - CEA/DSM - CNRS - Universit\'{e} Paris Diderot, B\^{a}t. 709, CEA-Saclay, F-91191 Gif-sur-Yvette Cedex, France\goodbreak
\and
Laboratoire Traitement et Communication de l'Information, CNRS (UMR 5141) and T\'{e}l\'{e}com ParisTech, 46 rue Barrault F-75634 Paris Cedex 13, France\goodbreak
\and
Laboratoire de Physique Subatomique et Cosmologie, Universit\'{e} Grenoble-Alpes, CNRS/IN2P3, 53, rue des Martyrs, 38026 Grenoble Cedex, France\goodbreak
\and
Laboratoire de Physique Th\'{e}orique, Universit\'{e} Paris-Sud 11 \& CNRS, B\^{a}timent 210, 91405 Orsay, France\goodbreak
\and
Lawrence Berkeley National Laboratory, Berkeley, California, U.S.A.\goodbreak
\and
Lebedev Physical Institute of the Russian Academy of Sciences, Astro Space Centre, 84/32 Profsoyuznaya st., Moscow, GSP-7, 117997, Russia\goodbreak
\and
Max-Planck-Institut f\"{u}r Astrophysik, Karl-Schwarzschild-Str. 1, 85741 Garching, Germany\goodbreak
\and
National University of Ireland, Department of Experimental Physics, Maynooth, Co. Kildare, Ireland\goodbreak
\and
Nicolaus Copernicus Astronomical Center, Bartycka 18, 00-716 Warsaw, Poland\goodbreak
\and
Niels Bohr Institute, Blegdamsvej 17, Copenhagen, Denmark\goodbreak
\and
Niels Bohr Institute, Copenhagen University, Blegdamsvej 17, Copenhagen, Denmark\goodbreak
\and
Optical Science Laboratory, University College London, Gower Street, London, U.K.\goodbreak
\and
SISSA, Astrophysics Sector, via Bonomea 265, 34136, Trieste, Italy\goodbreak
\and
School of Physics and Astronomy, Cardiff University, Queens Buildings, The Parade, Cardiff, CF24 3AA, U.K.\goodbreak
\and
Sorbonne Universit\'{e}-UPMC, UMR7095, Institut d'Astrophysique de Paris, 98 bis Boulevard Arago, F-75014, Paris, France\goodbreak
\and
Space Research Institute (IKI), Russian Academy of Sciences, Profsoyuznaya Str, 84/32, Moscow, 117997, Russia\goodbreak
\and
Space Sciences Laboratory, University of California, Berkeley, California, U.S.A.\goodbreak
\and
Special Astrophysical Observatory, Russian Academy of Sciences, Nizhnij Arkhyz, Zelenchukskiy region, Karachai-Cherkessian Republic, 369167, Russia\goodbreak
\and
Sub-Department of Astrophysics, University of Oxford, Keble Road, Oxford OX1 3RH, U.K.\goodbreak
\and
The Oskar Klein Centre for Cosmoparticle Physics, Department of Physics,Stockholm University, AlbaNova, SE-106 91 Stockholm, Sweden\goodbreak
\and
UPMC Univ Paris 06, UMR7095, 98 bis Boulevard Arago, F-75014, Paris, France\goodbreak
\and
Universit\"{a}t Heidelberg, Institut f\"{u}r Theoretische Astrophysik, Philosophenweg 12, 69120 Heidelberg, Germany\goodbreak
\and
Universit\'{e} de Toulouse, UPS-OMP, IRAP, F-31028 Toulouse cedex 4, France\goodbreak
\and
University Observatory, Ludwig Maximilian University of Munich, Scheinerstrasse 1, 81679 Munich, Germany\goodbreak
\and
University of Granada, Departamento de F\'{\i}sica Te\'{o}rica y del Cosmos, Facultad de Ciencias, Granada, Spain\goodbreak
\and
University of Granada, Instituto Carlos I de F\'{\i}sica Te\'{o}rica y Computacional, Granada, Spain\goodbreak
% ...
}

  \abstract
    {The Virgo cluster is the largest Sunyaev-Zeldovich (SZ) source in the sky, both in terms of angular size and 
    total integrated flux.  
    \Planck's wide angular scale and frequency coverage, together with its high sensitivity, allow a detailed study of this large object through the SZ effect.  
    Virgo is well resolved by \Planck, showing an elongated structure, which correlates well with the morphology 
    observed from X-rays, but extends beyond the observed X-ray signal.
    We find a good agreement between the SZ signal (or Compton paranmeter, $y_{\rm c}$) observed by \Planck\ and the expected signal inferred 
    from X-ray observations and simple analytical models. 
    Due to its proximity to us, the gas beyond the virial radius can be studied with unprecedented sensitivity by 
    integrating the SZ signal over tens of square degrees.  
    We study the signal in the outskirts of Virgo and compare it with analytical models and 
    a constrained simulation of the environment of Virgo.
    \Planck\ data suggest that significant amounts of low-density plasma surround Virgo out to 
    twice the virial radius. 
    We find the SZ signal in the outskirts of Virgo to be consistent with a simple model that extrapolates 
    the inferred pressure at lower radii while assuming that the temperature stays in the keV range 
    beyond the virial radius. The observed signal is also consistent with simulations and points to a shallow 
    pressure profile in the outskirts of the cluster. This reservoir of gas at large radii can be linked 
    with the hottest phase of the elusive warm/hot intergalactic medium. 
    Taking the lack of symmetry of Virgo into account, we find that a prolate model is favoured by the combination 
    of SZ and X-ray data, in agreement with predictions.
    Finally, based on the combination of the same SZ and X-ray data, we constrain the total amount of gas in Virgo.
    Under the hypothesis that the abundance of baryons in Virgo is representative of the cosmic average, 
    we also infer a distance for Virgo of approximately 18\,Mpc, in good agreement with 
    previous estimates.}


   \maketitle
%
%________________________________________________________________
%%%%%%%%%%%%%%%%%%%%%%%%%%%%%%%%%%%%%%%%%%%%%
\section{Introduction}\label{introduction}
%%%%%%%%%%%%%%%%%%%%%%%%%%%%%%%%%%%%%%%%%%%%%

The Virgo cluster is the closest cluster to our Galaxy (smaller groups like Leo are closer but are significantly 
less massive). 
Virgo is a moderate cluster in terms of its mass, $M \approx (4$--$8) \times10^{14} {\rm M}_{\odot}$ 
\citep{Vaucouleurs1960,Hoffman1980,Bohringer1994,Karachentsev2010,Karachentsev2014}, but 
it is the largest cluster in terms of angular size (tens of square degrees). The combination of moderate mass and large angular size makes this cluster the largest single source for the Sunyaev-Zeldovich (SZ) effect in the sky in terms of integrated flux \citep{Taylor2003}. The Coma cluster on the other hand, is the most prominent cluster in terms of signal-to-noise \citep{PlanckComa} but subtends a significantly smnaller solid angle than Virgo.   

Despite its large total flux, the strength of the SZ signal 
(or more specifically, the surface brightness) is expected to be low for this moderate mass 
cluster, as shown by \cite{Diego2008}, making it a challenging object to study. 

Due to its proximity to us \citep[distance between 15\,Mpc and 21\,Mpc][] 
{Sandage1976,Freedman1994,Pierce1994,Federspiel1998,Graham1999,Tammann2000,Jerjen2004,Mei2007}, 
Virgo presents a unique opportunity to use 
\Planck\footnote{\Planck\ (\url{http://www.esa.int/Planck}) is a project of the 
European Space Agency (ESA) with instruments provided by two scientific 
consortia funded by ESA member states and led by Principal Investigators 
from France and Italy, telescope reflectors provided through a collaboration 
between ESA and a scientific consortium led and funded by Denmark, and 
additional contributions from NASA (USA).} 
to study the SZ effect in the outskirts of clusters with \Planck.

The angular size of the virial radius, $R_{\rm vir}$, of Virgo is
$4\deg$ (for consistency we adopt the same virial radius, $R_{\rm vir}$ or $R_{200}$, as \citealt{Urban2011}) 
corresponding to 1\,Mpc (for a distance of $D=15\,\mathrm{Mpc}$). 
The area enclosed between  $R_{\rm vir}$ and $1.25 R_{\rm vir}$ is an impressive 
$30$ deg$^2$. No other cluster offers such a large area, enabling us to average the SZ signal 
and increase the sensitivity to an unprecedented level. 
Hence, the Virgo cluster offers a unique opportunity to study the elusive warm/hot 
intergalactic medium (or WHIM), which is believed to exist around clusters and has, so far, evaded a clear detection. 
The WHIM gas is expected to have temperatures of $10^5$--$10^7\,\mathrm{K}$ and densities of $10$--$30$ times the 
mean density of the Universe \citep{Dave2001}. 
The strength of the SZ signal around Virgo due to the WHIM is expected to be of the order of a microkelvin 
at the relevant \Planck\ frequencies as shown by \cite{Diego2008}. 
This level of sensitivity in the outskirts of clusters can be reached with \Planck\ only after 
averaging large areas of the sky. This in turn is possible only in the Virgo cluster, which subtends such a large solid angle 
degrees while other clusters subtend areas of the order of 1 deg$^2$  
at most (with Coma a little above this number). 
The evidence of ram pressure acting in the outskirts of Virgo \citep{Kenney2004} also supports 
the hypothesis that vast amounts of gas are still present at these distances from the centre and hence 
possibly detectable by \Planck.

%__________________________________________________________________
\begin{figure}  % MADE BY:  (loki) Virgo_Planck/AuxPro/Plot_VirgoComa_Klaus.pro
   \centering
   \includegraphics[width=9.0cm]{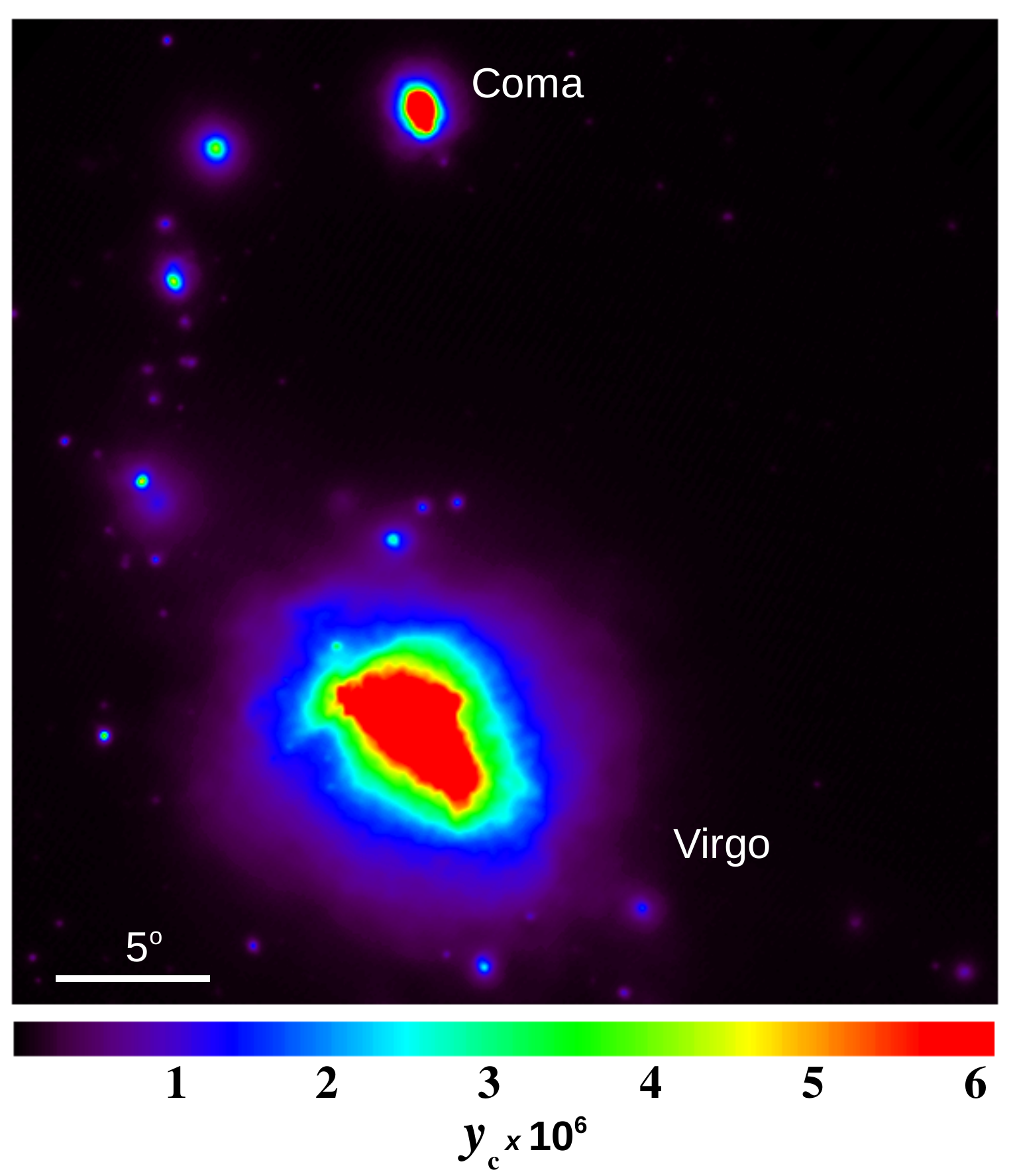}
      \caption{Simulated Virgo cluster from a constrained simulation in units of $10^{-6}$ 
               times the Compton parameter \citep{Dolag2005}. 
               To increase the contrast, the Compton parameter beyond $6\times 10^{-6}$ has been saturated. 
               The brightest cluster in the top is Coma. In this and other figures 
               (unless otherwise noted) the coordinates are Galactic, the projection is orthographic, and 
               the field of view is 31\pdeg94 across or 8.3 (11.9)\,Mpc assuming Virgo 
               is at a distance of 15 (21.4)\,Mpc. The North Galactic pole is up and Virgo's centre (M87) is $4^{\circ}$ south 
               of the centre of the images.}
         \label{Fig_Sim_Virgo}
   \end{figure}
%__________________________________________________________________

%__________________________________________________________________
\begin{figure*} % MADE BY : plot_Tau_T_Beta.pro
   \centering
   \includegraphics[width=18.0cm]{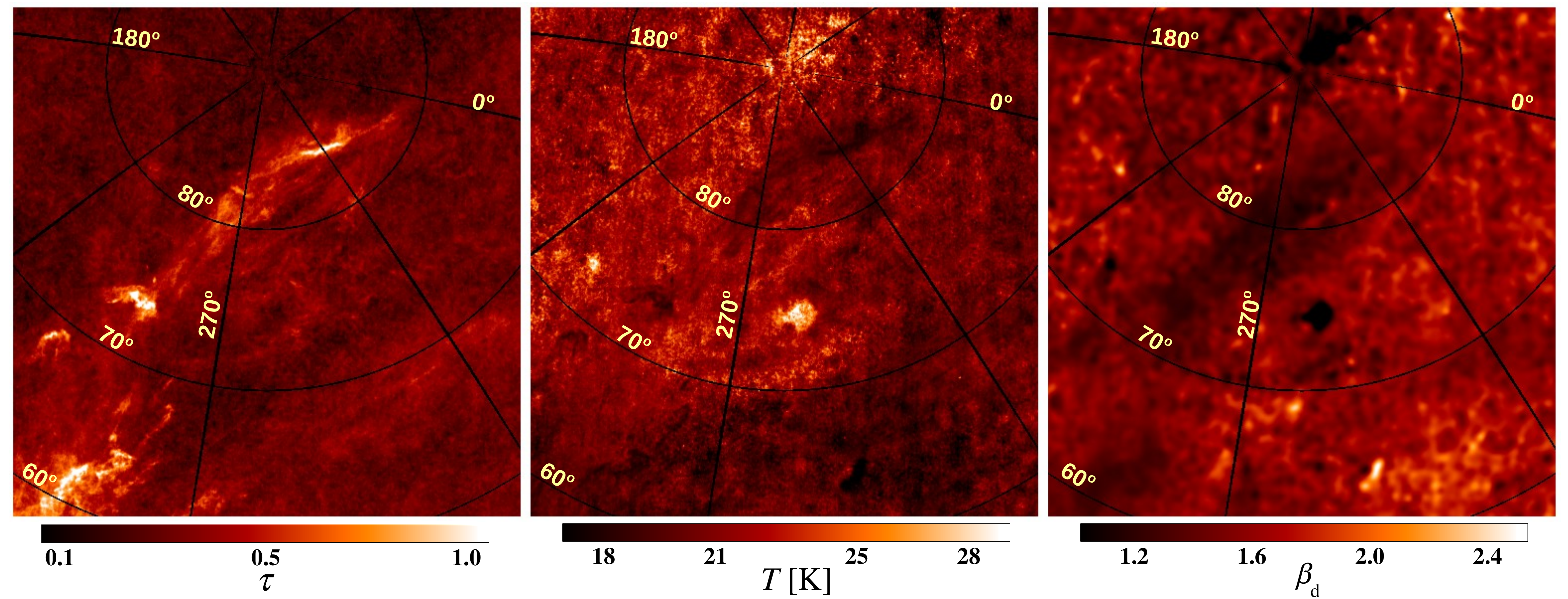}
      \caption{\emph{Left} to \emph{right}: Optical depth $\tau$, dust temperature $T$ (in kelvin), and spectral 
               index $\beta_{\rm d}$ of the dust component in the Virgo region. Some clusters in this field of view (Coma to the North, Leo to the 
               East, and Virgo near the centre) can be clearly spotted as artefacts in the $T$ and 
               $\beta_{\rm d}$ maps. The coordinates are Galactic and the field of view 
               resembles that of Fig.~\ref{Fig_Sim_Virgo} ($31\pdeg9$) and other figures in this paper.}
         \label{Fig_Dust}
   \end{figure*}
%__________________________________________________________________

The full-sky coverage and wide frequency range of \Planck\ makes it an ideal instrument to study the SZ effect 
in large objects like Virgo \citep{2011A&A...536A...8P,2011A&A...536A...9P,2011A&A...536A..10P}. 
\Planck\ performs better than ground-based experiments when studying diffuse SZ signals that extend over 
relatively large angular scales. Ground-based observations at large angular scales are affected by atmospheric 
fluctuations, which need to be removed, removal of which distorts the modes that include the signal on large 
angular scales. \Planck\ data do not suffer from these limitations, and hence diffuse, low-surface brightness objects 
can be resolved and detected by \Planck\ provided that their solid angle is large enough. The main limitations of \Planck\ 
data are Galactic foregrounds and the cosmic microwave background (CMB) signal; however, the frequency coverage 
and sensitivity of \Planck\ allow for the detailed foreground and CMB removal that is necessary to study this cluster. 

The proximity of Virgo to us also enables detailed studies in other bands. 
Of particular interest are the results derived from X-rays, since they trace the same plasma 
that is responsible for the SZ effect. 
Virgo is also the brightest cluster in terms of integrated X-ray flux. It has been 
studied in detail by different X-ray observatories (\ROSAT, \XMM, and \CHANDRA), although only \ROSAT, with 
its full-sky coverage, provides a complete picture of Virgo and its surroundings. 
On the other hand, the superior resolution of \XMM\ and \CHANDRA\ has allowed more detailed studies of the 
inner structure of Virgo. In particular, \cite{Ghizzardi2004} (G04 hereafter) present an analysis of the 
central 100\,kpc region around M87 (with \CHANDRA\ and \XMM\ X-ray data). Using \XMM\ data,  
\citep[][U11 hereafter]{Urban2011} study the X-ray emission in Virgo 
out to the virial radius ($R_{\rm vir} \approx 1$\,Mpc), but only through a narrow sector of the Virgo cluster. 

When combined with X-ray data, the SZ signal from \Planck\ has the potential to discriminate between different models that are indistinguishable with X-ray data alone. 
This fact is exploited in this paper, as one of its main objectives, 
to infer properties of the distribution of the plasma in this cluster. 
Also, the possibility to resolve Virgo with \Planck\ and hence to study 
different regions independently allows for a more detailed study of these regions. 
Due to the different dependencies of the SZ effect and X-ray emission on the electron density, 
resolved images also allow us to test possible deviations from a smooth distribution of the plasma.
 
Although the increased angular resolution provided by the recent X-ray observatories, in particular \CHANDRA, 
has constrained the amount of clumpiness, it is interesting to study possible differences between the predicted 
SZ effect based on X-ray-derived models and the SZ signal observed by \Planck. 
The relative proximity of Virgo means that at the resolution of \Planck\ any features larger than 26\,kpc 
can potentially be resolved. This is an unprecedented resolution for galaxy cluster 
studies. As opposed to other typical clusters, the geometry of Virgo 
reveals a complex structure with different clumps being separated by up to several megaparsecs \citep{Jerjen2004}.

At the centre of Virgo is M87, one of the most powerful known AGN, flanked by two very luminous radio jets. 
The radio emission marks the regions where large amounts of energy are being injected into the cluster medium. 
In X-rays, cold fronts and shock waves have been identified that are also linked to the extraordinary activity of the 
supermassive black hole at the centre of M87. Perhaps the clearest evidence of the extreme activity 
surrounding M87 is the presence of radio lobes with a geometry that suggests sub-sonic injection of colder gas 
into the intracluster medium \citep{Churazov2001,Forman2007}. 
The activity of the AGN has an impact on the 
X-ray emissivity of the surrounding gas at supergalactic scales. 

As a theoretical counterpart, we use the Virgo cluster from a
constrained simulation of the local Universe. We adopt the initial
conditions from \cite{Mathis2002},  
who perform $N$-body numerical simulations and demonstrated that the evolved state of these initial
conditions provides a good match to the large-scale structure observed
in the local Universe. Here, many of the most prominent nearby
clusters (among them Virgo and Coma) can be identified directly with
halos in the simulations, with a good agreement for sky position and
virial mass. \cite{Dolag2005} repeated these simulations including
also the baryonic component, demonstrating that the simulated SZ
signal reproduces the expected SZ signal. In this work we use a new
simulation that includes a wealth of physical processes known to be
important to reproduce realistic galaxy clusters, most importantly
the feedback from star formation and from super-massive
black holes. We use an implementation that also matches various
observed properties of AGN, as described in the work of \cite{Hirschmann2014},
where details of the implemented physical processes can be
found.

This physical treatment within the simulations results in
averaged pressure profiles, which match well the observed ones from
\Planck, as demonstrated in \cite{planck2012-V}. However, one has to
keep in mind that the uncertainties in the construction of the initial
conditions are still substantial, so that the details of the Virgo
cluster formed in the simulations will be different from the real,
observed one. Especially given that the current state of the AGN in the centre of
the simulated Virgo cluster is not constrained at all, we therefore
expect large differences particularly in the central part.
Figure~\ref{Fig_Sim_Virgo} shows the region of the sky that contains the
simulated Virgo and Coma clusters in units of the Compton parameter,
$y_{\rm c}$, 
\begin{equation}
y_{\rm c} = \frac{\sigma_{\rm T}\, k_{\rm B}}{m_{\rm e} \, c^2} \int n_{\rm e} \, T \, dl
\end{equation}
where the integral is along the line of sight and
$\sigma_{\rm T}$, $k_{\rm B}$, $m_{\rm e}$, $n_{\rm e}$, and $T$ are the Thomson cross section,
Boltzmann constant, mass of the electron, electron density, and
electron temperature, respectively.

%__________________________________________________________________
\begin{figure}  % MADE BY: Plot_DiffMaps.pro
   \centering
   \includegraphics[width=9cm]{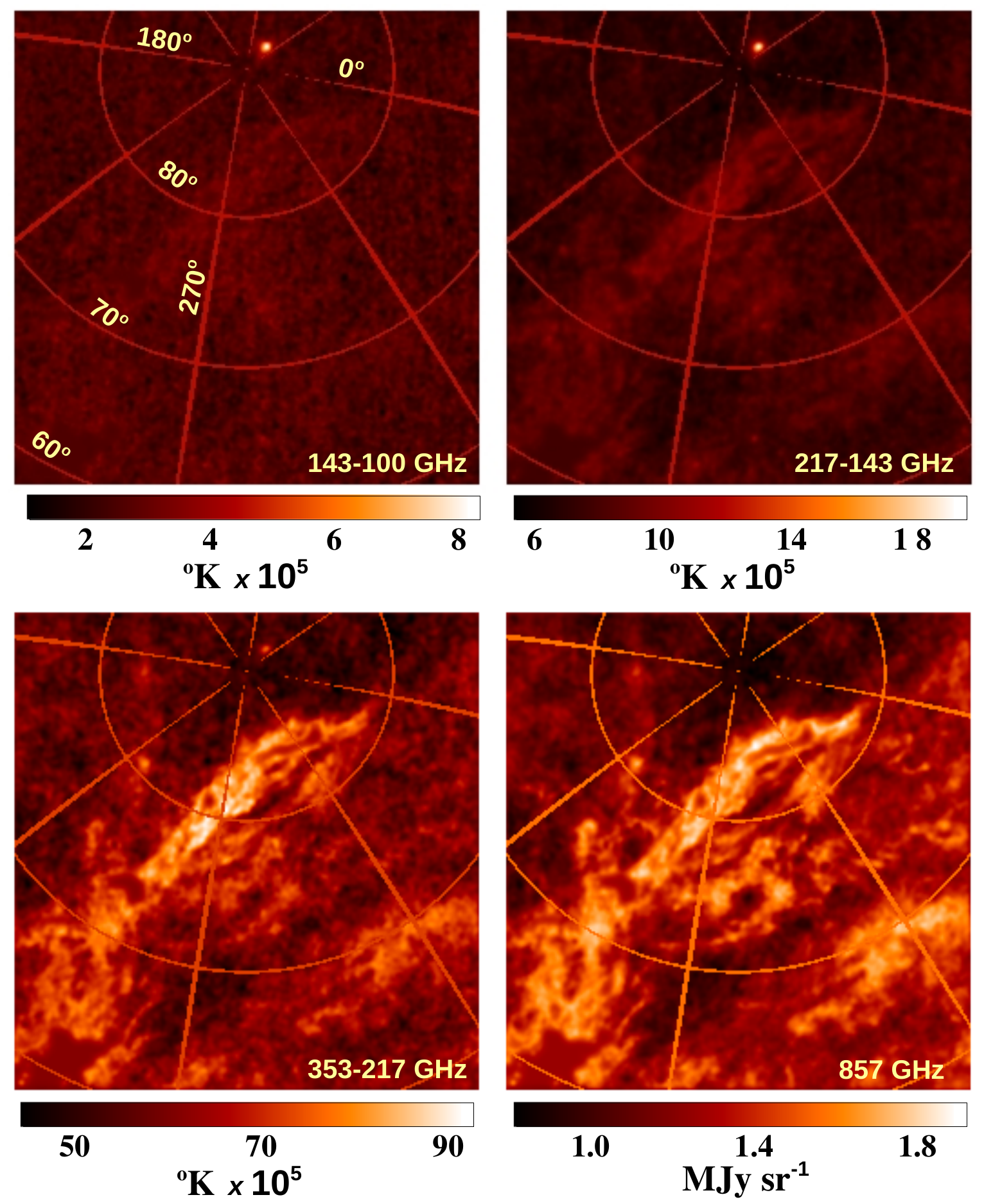}
      \caption{Differences (in kelvin) between the individual \Planck\   bands. The lower right panel 
               shows the original data at 857\,GHz (in MJy sr$^{-1}$). 
               All maps have been masked and degraded to the same 
               resolution before calculating the differences. In all differences the SZ effect should 
               appear as a positive signal. The Coma cluster is clearly visible in the top of the images 
               of the low-frequency differences.}
         \label{Fig_MapDiffs}
\end{figure}
%__________________________________________________________________

The positions of Virgo and other simulated clusters in the field of
view in Fig.~\ref{Fig_Sim_Virgo} are approximately the same as the actual
ones in the real world, as seen in previous studies \citep{Mathis2002,
Dolag2005}. The image shows the Compton parameter, $y_{\rm c}$, and
has been saturated above $y_{\rm c} = 6 \times 10^{-6}$ for contrast purposes.
When compared with Coma (the highest signal-to-noise (S/N) SZ source in
the sky), it is evident that despite its low surface brightness, Virgo
is expected to be the brightest source in terms of integrated signal.
Also from the same figure, it is evident that the outskirts of the
Virgo cluster cover a significant area of many square degrees, where
the elusive WHIM could be detected after integrating over this large
area. Some evidence of its existence already exists from observations
\citep[e.g.,][]{Yoon2012}.

The resolution of \Planck\ does not allow us to study the pressure profile of clusters in detail, since most of them 
appear unresolved or barely resolved. A few exceptions, like the Coma cluster, have allowed a study of the 
pressure profile of clusters in more detail \citep{PlanckComa}. 
However, Virgo is a significantly less massive cluster (for which SZ effect observations are rare) 
and offers a unique opportunity to study the pressure profile in a moderate-mass system, for which the constraints 
are poor. The pressure profile has also been studied with \Planck\ in a sample of 62 nearby massive clusters 
detected at high significance with \Planck\ and with good ancillary X-ray data \citep{PlanckPressure}.

%__________________________________________________________________
\begin{figure*}  % MADE BY: PlotSZ_070_100_143.pro
   \centering
   \includegraphics[width=18.5cm]{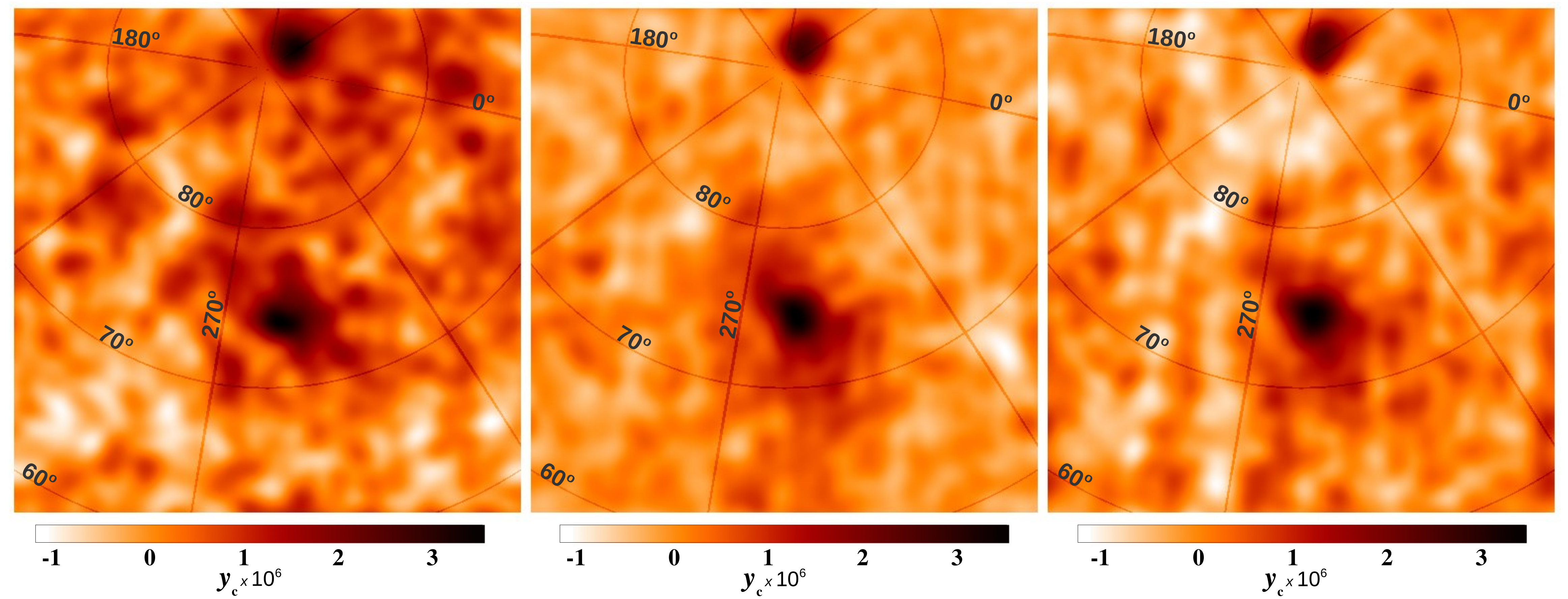}
      \caption{\emph{Left} to \emph{right}: individual SZ maps at 70, 100, and 143\,GHz in units of $10^6$ times the Compton parameter. 
               The field of view is $31\pdeg9$ and the coordinates are Galactic. 
               Virgo is the extended source below the centre. 
               The Coma cluster is also clearly visible in the top of the map. The maps have been smoothed 
               to a resolution of $1\pdeg5$ to highlight the faintest structures.}
         \label{Fig_SZ_070_100_143}
\end{figure*}
%__________________________________________________________________

Virgo allows us to extend the study of the pressure profile to the poorly explored regime 
of low to moderate-mass clusters and also to extend the distances at which the pressure can be 
observed. This constitutes a primary objective of the current paper. 

Given its relative proximity to us, a distance estimate based on the redshift of the galaxies within Virgo
can be inaccurate due to the peculiar velocities of those galaxies and/or our own movement 
towards Virgo. 
On the other hand, the distance to Virgo is large enough that distance measurements based on Cepheids or planetary 
nebulae are also difficult to make. 
As a consequence, the absolute distance of the Virgo cluster is still a matter of debate, with measurements from various 
techniques (Cepheids, globular clusters, planetary nebulae, surface brightness fluctuations, and supernovae) 
resulting in a range of distances for Virgo ranging from 15\,Mpc to 21\,Mpc  
\citep{Sandage1976,Freedman1994,Pierce1994,Federspiel1998,Graham1999,Tammann2000,Jerjen2004,Mei2007}. 
Part of the discrepancy may be due to the distribution of spiral galaxies in the field of view, which 
concentrate in an elongated region extending from 13 to 30\,Mpc, closely aligned with the line of sight. This affects 
the distance moduli estimated based on Tully-Fisher analyses \citep{Fukugita1993}. 
Following G04, and in order to keep consistency with that work, 
we adopt a default distance of 21\,Mpc for the computation of the X-ray count rate and the conversion 
from physical to angular scales. The implications of adopting a different distance are discussed in 
Sect.~\ref{subsect_dist}, where we also derive an independent distance estimate based on the combined 
SZ and X-ray observations. 

The paper is organized as follows. Section~\ref{Planck_Data} gives a brief description of the \Planck\ data used 
for this study. In Sect.~\ref{selection_2} we describe the data analysis and potential sources of systematics 
that could affect the extraction of the SZ signal from \Planck\ data. 
Section~\ref{sect_xrays} describes the X-ray \ROSAT\ observations of Virgo. In Sect.~\ref{sect_models} we model the SZ effect and X-ray emission from the cluster and discuss the models derived from \XMM\ and \CHANDRA. 
Section~\ref{Model_Results} presents results on standard radial profiles of the SZ and X-ray data and 
a comparison with different models.
In Sect.~\ref{Beyond} we focus our attention on the signal beyond the virial radius and constrain the distance to 
Virgo from the combination of SZ and X-ray data. 
Finally, we present the conclusions in Sect.~\ref{sect_conclusion}.

%_______________________SECTION \PLANCK\ DATA ___________________________________________

%%%%%%%%%%%%%%%%%%%%%%%%%%%%%%%%%%%%%%%%%%%%%%%%%%%%%%%%%%
\section{{\Planck} data}\label{Planck_Data}
%%%%%%%%%%%%%%%%%%%%%%%%%%%%%%%%%%%%%%%%%%%%%%%%%%%%%%%%%%
%=========================================================

%-----------------------------------------------------------------------
% To include all Planck 2015 Results papers in the reference lists
%-----------------------------------------------------------------------
%\alltwentyfifteenresultspapers

\Planck's sensitivity, angular resolution, and frequency coverage make it a
powerful instrument for Galactic and extragalactic astrophysics
as well as cosmology. An overview of the latest results from \Planck\ can be 
found in \cite{planck2014-a01}. 
This paper is based on \Planck's full survey mission, where the whole sky has been covered 
over 5 times by the high frequency instrument (HFI) and 8 times by the low frequency instrument (LFI). We refer to 
\cite{planck2014-a03}, \cite{planck2014-a04}, \cite{planck2014-a05}, \cite{planck2014-a06}, 
\cite{planck2014-a07}, \cite{planck2014-a08} and \cite{planck2014-a09} 
for technical information about the calibration, processing of the data, mapmaking, instrument response, 
and beams of the two instruments, LFI and HFI. 
For this particular work, we rely on the \Planck\ bands at 70, 100, 143, 217, and 353\,GHz. However, the 
bands at 545 and 857\,GHz are also used to test for Galactic residuals. 

From the full-sky \Planck\ maps, we extract square patches of $31\pdeg9 \times 31\pdeg9$. 
The centre of the field of view is chosen such that other prominent clusters (like nearby Coma) 
are also included in the field of view for comparison. 
Other smaller, but still prominent, clusters like the Leo cluster also fall in our field of view. 
We extract patches of \Planck's LFI and HFI data, as well as ancillary data (e.g., NVSS, IRAS, and \ROSAT). 
The ancillary data are used to check for possible systematic effects (contamination from radio or infrared point sources, 
Galactic contamination or Galactic dust variability) but also to use these data in combination with \Planck\   
(in particular \ROSAT). 
Patches from the ``half-ring'' differences and each individual survey (eight for LFI and five for HFI) are 
also extracted to study possible systematic effects and to understand the properties of the noise 
maps.

%_________________     S Z     I N     V I R G O     _________________________

%%%%%%%%%%%%%%%%%%%%%%%%%%%%%%%%%%%%%%%%%%%%%%%%%%%%%%%%%%
\section{Extracting the SZ effect in Virgo from {\Planck} data}\label{selection_2}
%%%%%%%%%%%%%%%%%%%%%%%%%%%%%%%%%%%%%%%%%%%%%%%%%%%%%%%%%%
%=========================================================
To isolate the SZ effect from Virgo, the dominant signals from the CMB and the Galaxy need to be removed. 
As shown in Fig.~\ref{Fig_Dust}, the Galactic emission around the Virgo cluster 
exhibits a complex pattern, with significant variations in the optical depth $\tau$, dust temperature $T_{\rm d}$, 
and spectral index (of a modified blackbody model) $\beta_{\rm d}$. In the optically thin limit, the specific intensity 
$I_{\nu}$ from a population of grains is well approximated by a modified blackbody,
\begin{equation}
I_{\nu} \propto \tau \, \nu^{\,\beta_{\rm d}} \, B(\nu,T_{\rm d}),
\end{equation}
where $B(\nu,T_{\rm d})$ is the Planck blackbody function. 
From Fig.~\ref{Fig_Dust} the dust temperature ranges from 18 to 28\,K,  
while the emissivity index, $\beta_{\rm d}$, runs from 1 to 2.4. 
The temperature and spectral index are derived from the combination of the 353\,GHz, 545\,GHz, 857 
GHZ \Planck\ bands and the 100\,$\mu$m IRAS band, which are fit to the form of Eq. (2)  
after subtraction of an average (constant) estimate of the background (CIB) levels and the CMB \citep[see][]{Planck2013_XI}.
This figure summarizes some of the main challenges in our analysis arising from the complexity of the 
Galactic contamination around Virgo. 
There is a distinctive feature in the north-west with a lower temperature and higher 
spectral index, crossing relatively close to the northern part of Virgo. 
A ring of diffuse Galactic dust also surrounds the centre of the Virgo region. 
In addition, although not evident in these maps, 
the zodiacal light extends horizontally in the bottom part of our field of view, although most of its contribution 
can be removed together with the Galaxy, since its spectral behaviour is very similar to Galactic emission. 
The 217\,GHz band, critical for the removal of the CMB, is still significantly contaminated 
by Galactic dust (seen clearly, for instance, in the difference between the 217\,GHz and 100\,GHz maps). 
In standard SZ maps produced via an internal linear combination (ILC) it is generally assumed that  
there is no spatial variability in the dust spectral index or the temperature, so some residual 
Galactic contamination is expected to remain in the SZ maps. 
Instead of producing an ILC SZ map, the temperature and spectral index 
information could, in principle, be used to overcome this problem by taking their spatial 
variability into account.
However, the SZ signal itself contaminates the estimates of the dust temperature and spectral index 
at the positions of clusters, resulting in high temperature and low spectral 
index, as shown in Fig.~\ref{Fig_Dust} at the positions of Coma and Virgo. 
A third cluster, Leo, is also clearly seen to the East in the temperature and spectral index maps.

%__________________________________________________________________
\begin{figure}  % (loki) MyPapers/Virgo/FinalVersion/VirgoSZGrid.odp or Virgo_Planck/VirgoSZGrid.odp
   \centering
   \includegraphics[width=9cm]{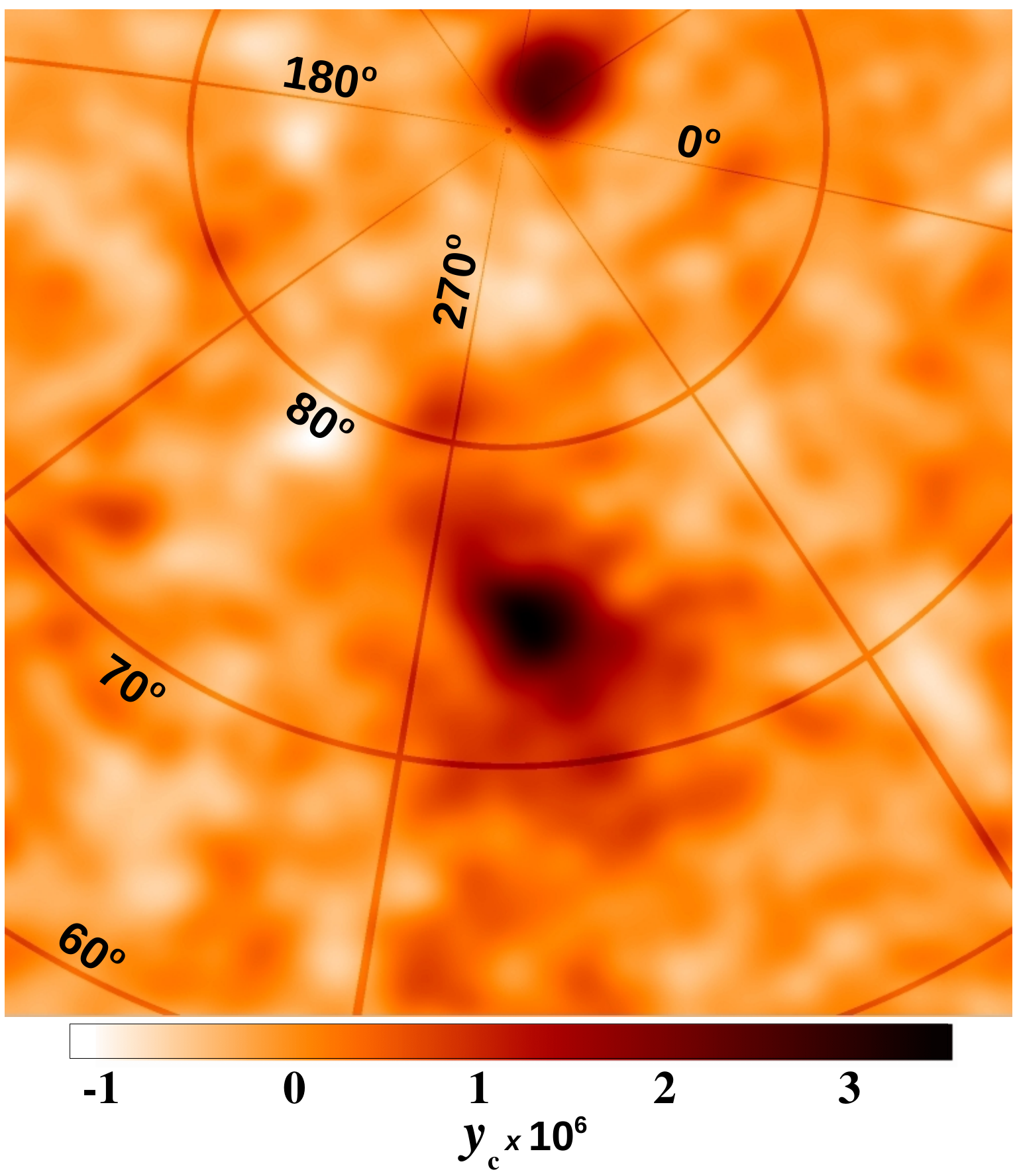}
      \caption{Virgo SZ map showing its position and extent in Galactic coordinates. 
               The Compton parameter $y_{\rm c}$ is plotetd here.
               The resolution is $1\pdeg5$.}
         \label{Fig_SZ_Grid}
\end{figure}
%__________________________________________________________________

\subsection{Internal Linear Combination for Virgo}
%%%%%%%%%%%%%%%%%%%%%%%%%%%%%%%%%%%%%%%%%%%%%%%%%%%
The Virgo cluster is clearly seen in the latest \Planck\  maps of the thermal SZ effect \citep{planckSZallsky}. 
Although 
these maps provide a high-quality all-sky vision of the SZ effect at all scales, certain regions in the sky are still contaminated by 
Galactic residuals. These could be reduced by performing a specific component separation in that region. As discussed in section 
\ref{selection_2}, due to the Galactic contamination, we use a specific and tailored ILC method 
to produce the SZ map for Virgo. 
First, we use only the \Planck\ maps that contain the 
least amount of Galactic contamination. In particular, in our analysis we do not use 
the maps at 545 and 857\,GHz, frequently used in other ILC methods in \Planck\, e.g in {\it MILCA} \citep{Hurier2013} 
and {\it NILC} \citep{Delabrouille2009,Remazeilles2011}, as tracers of the Galactic dust. 
Due to variations in the dust spectral index and temperature, these maps introduce spurious residuals when 
combined with (or extrapolated to) low frequencies. 
Second, we avoid filtering of the large scales in the maps. The filtering is 
normally used in standard ILC maps like {\it MILCA} \citep{Hurier2013} and {\it NILC} \citep{Delabrouille2009,Remazeilles2011}  
to reduce the negative impact of large-scale Galactic residuals.

%__________________________________________________________________
\begin{figure}  % MADE BY: Destripe_SZILC_DX11.pro + Correlation_SZ_Dust_DX11.pro + PlotVirgoSZDX11_Contours.pro 
   \centering 
   \includegraphics[width=9cm]{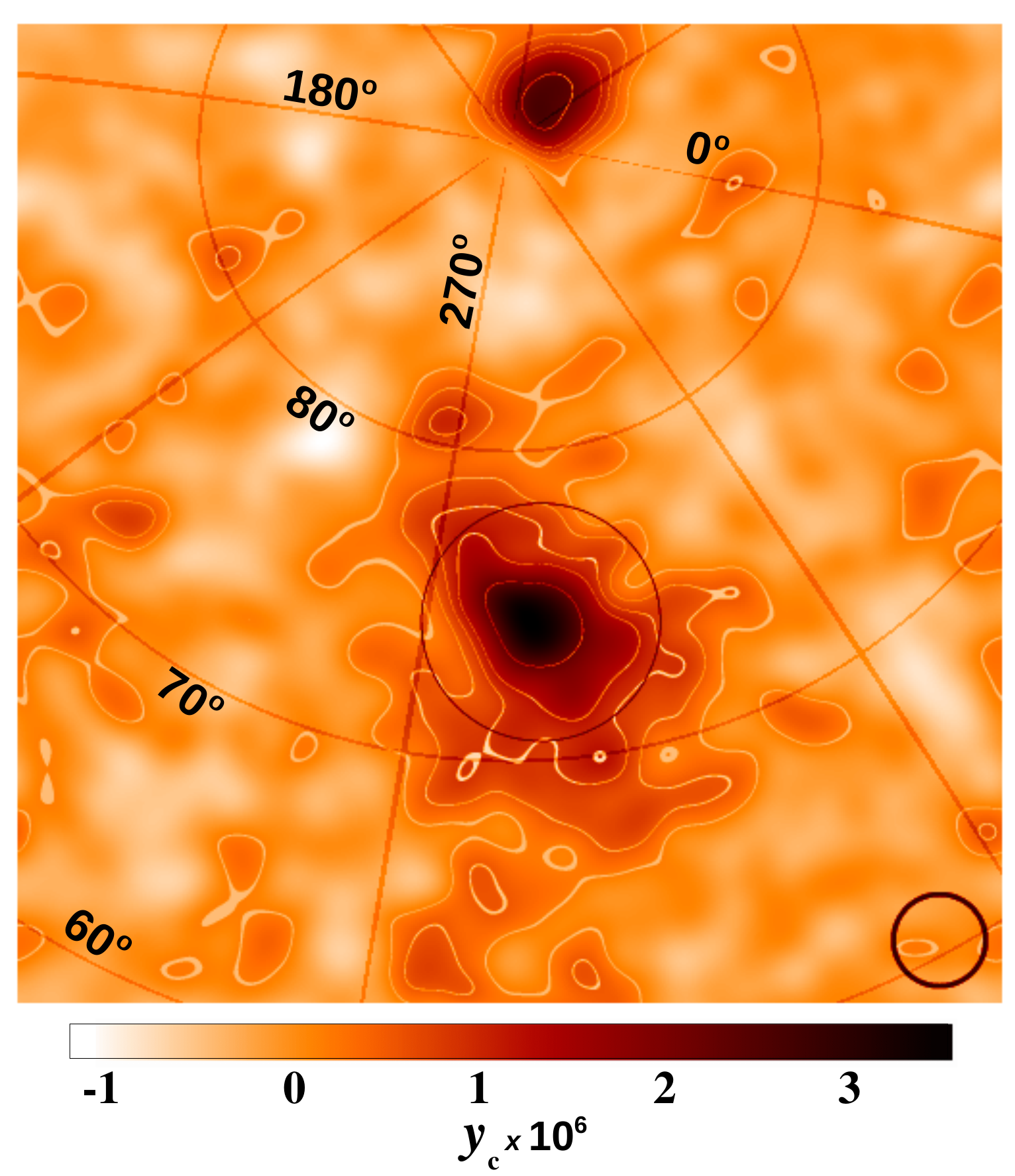}
      \caption{SZ effect map obtained after combining the individual maps in 
               Fig.~\ref{Fig_SZ_070_100_143} and after 
               reducing the impact of stripes (see Appendix~\ref{sect_stripes} below). 
               The colour scale is the same as in Fig.~\ref{Fig_SZ_Grid}. The contours are proportional to 1, 2, 3, 4, and 7 times the 
               dispersion of the background. 
               We only show the contours for the positive (negative in the map) 
               fluctuations of the Compton parameter. 
               Virgo's centre (M87) is $4\deg$ south of the centre of the image and the Coma cluster is 
               clearly seen at the top (i.e north). Other clusters like the Leo cluster (or group) 
               can also be appreciated near the east edge of the image (in the middle). The larger black 
               circle around Virgo marks the area enclosed within the virial radius ($3\pdeg9$). The small 
               circle at the bottom right has a diameter equal to the 
               FWHM of the Gaussian kernel used to smooth the image ($1\pdeg5$). The small circle 
               towards the north-east from Virgo marks the position of a feature in the SZ map 
               that is possibly due to Galactic contamination. The field of view is $31\pdeg9$. }
         \label{Fig_SZ_Contours}
\end{figure}
%__________________________________________________________________

%__________________________________________________________________
\begin{figure} % MADE BY: PlotVirgoSZDX11_Contours.pro and Correlation_SZ_Dust_DX11.pro
   \centering
   \includegraphics[width=9.0cm]{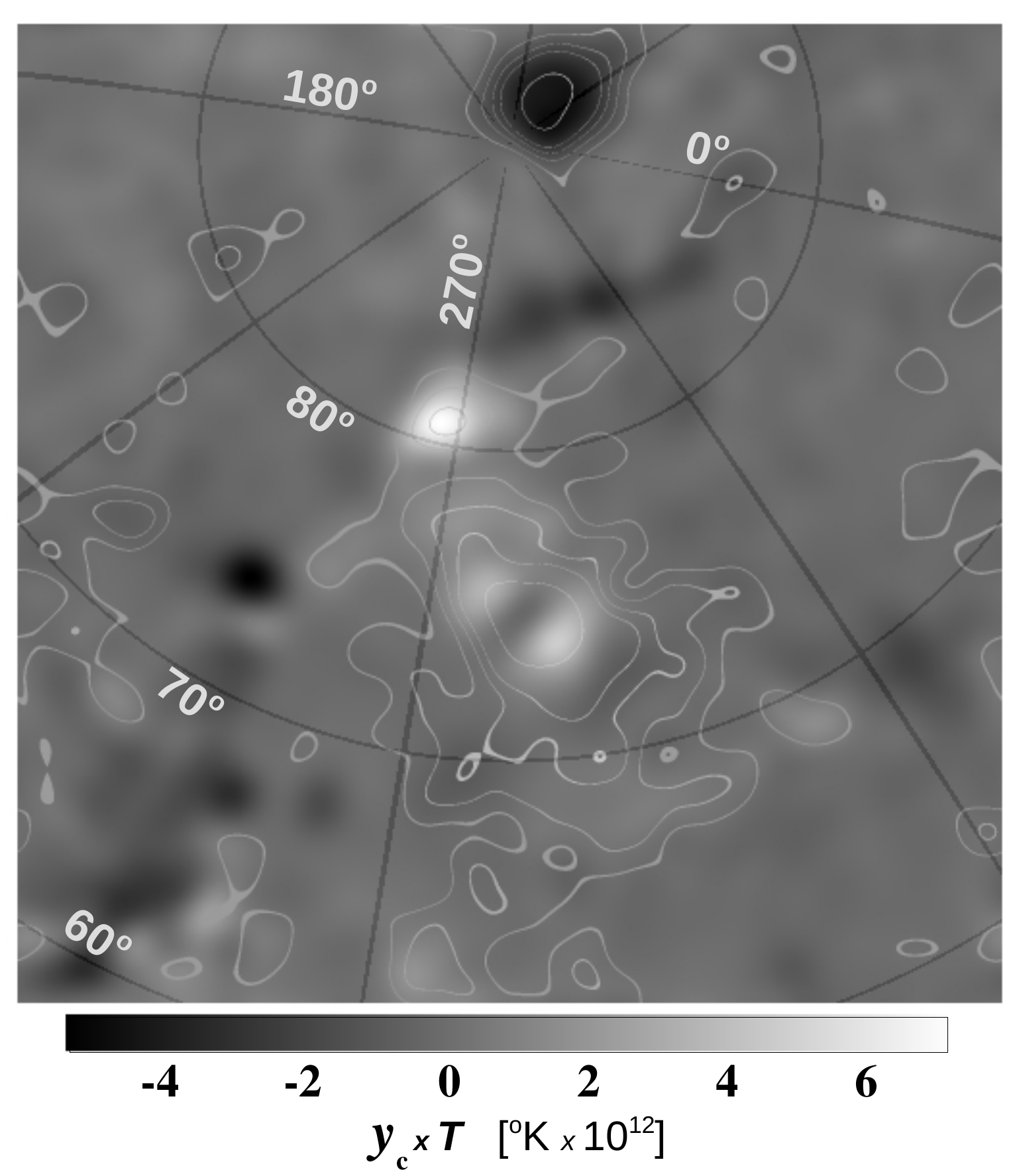}
      \caption{Map showing the correlation between the SZ map and the dust map. The 
               brighter areas show the regions where the correlation between the two maps is positive, 
               while the darker areas show areas with anti-correlation. Overlaid are the same 
               SZ effect contours as in Fig.~\ref{Fig_SZ_Contours}. The units are arbitrary and re-scaled by a factor of $10^{12}$, 
               but correspond to the dimensionless Compton parameter times temperature.}
         \label{Fig_SZ_Dust_correl}
   \end{figure}
%__________________________________________________________________

Since Virgo is a very extended source spanning many degrees, this filtering process could remove some 
of the large-scale features of Virgo.  
Instead, we produce individual SZ effect maps (at 70, 100, and 143\,GHz) by removing an estimate of the CMB, 
an estimate of the Galactic dust, and masking residual point sources. 
The estimate of the CMB is defined as a Galaxy-clean version 
of the 217\,GHz map, as described below. The estimate of the Galactic dust is obtained from simple differences 
of frequency bands (also described in more detail below). 
No model or template is assumed for either the Galaxy or the CMB and the SZ effect map is 
the result of a pure combination of \Planck\ maps (ILC) without any filtering in Fourier space (other than the 
degradation of the different maps to a common resolution). 
To derive the SZ effect map we use the \Planck\ data at 70, 100, 143, 217, and 353\,GHz.

%__________________________________________________________________
\begin{figure*} % MADE BY : PlotVirgoDX11_UrbanPointings.pro
   \centering
   \includegraphics[width=18.5cm]{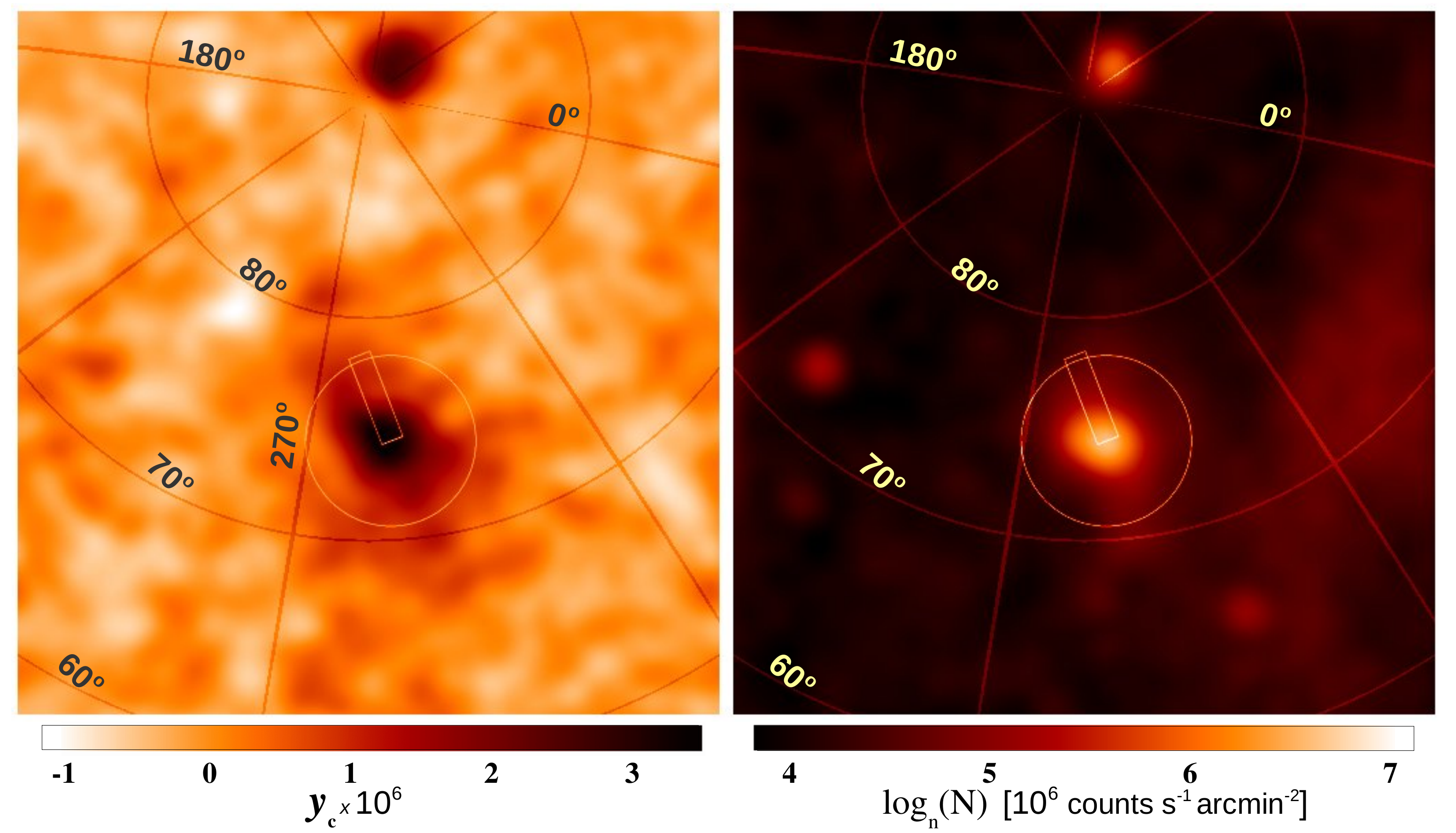}
      \caption{\emph{Left}: SZ map (stripe-reduced version) in units of the Compton parameter. \emph{Right}: \ROSAT\ All Sky Survey map in 
               Logarithmic scale (in units of $10^6$ counts\,s$^{-1}$\,arcmin$^{-2}$). 
               We indicate the sector (rectangular region) where \XMM\   
               pointings were used by \cite{Urban2011} to derive the electron density and temperature up to the 
               virial radius (circle).}
         \label{Fig_Planck_vs_RASS}
   \end{figure*}
%__________________________________________________________________

The band at 217\,GHz is particularly interesting because the thermal SZ is nearly zero in this band and 
it is dominated by CMB and Galactic foregrounds, our main sources of contamination for the SZ signal. 
All maps are degraded to a common resolution of 15\arcm\ (FWHM) so the final SZ effect map is obtained also 
with a resolution of 15\arcm. This resolution is typically poorer than that used in 
previous analyses of the SZ effect in \Planck\ data, where the typical resolution of the SZ effect map is between 
5\arcm\ and 10\arcm. 
In the particular case of Virgo, given its large angular size, degrading the maps to 15\arcm\ allows for the 
inclusion of the 70\,GHz band, which is relevant for SZ effect studies.
Our final SZ effect map is obtained as a weighted combination of the individual SZ effect maps obtained 
at 70, 100, and 143\,GHz.

More precisely, the cleaning process (i.e., the removal of the CMB and the Galaxy) is divided into three steps. 
\begin{enumerate}
    \item A template for the Galaxy is obtained from our maps and this template is used to clean 
the 217\,GHz band from Galactic residual and produce a {\it clean} map at 217\,GHz.
    \item The clean 217\,GHz band is taken as an estimate of the 
CMB that is subtracted from the other maps (70, 100, and 143\,GHz), producing individual estimates of the SZ effect 
that still contain Galactic residuals (plus instrumental noise 
and point source residuals) at the different frequencies. 
These Galactic residuals are reduced by using another alternative dust template, resulting in the three final SZ maps at 70, 
100, and 143\,GHz.

    \item The three SZ maps are combined to produce a single SZ map in units of the Compton parameter. 
\end{enumerate}

The final SZ map is dominated by SZ signal, but also contains instrumental noise, stripe structures due to 
baseline removal, and Galactic and extragalactic residuals. In summary, the SZ effect map at a given frequency 
$\nu$ (with $\nu=$ 70, 100, or 143\,GHz), is given by 
\begin{equation}
\mathrm{SZ}_{\nu} = M_{\nu} - M_{217}^\mathrm{c} - C_{\nu} \, D_\mathrm{T},
\label{eq_SZ effect}
\end{equation}
where $M_{\nu}$ is the \Planck\ map at frequency $\nu$,  $M_{217}^\mathrm{c}$ is the dust-cleaned 217\,GHz \Planck\   
map containing CMB, kinetic SZ effect, thermal SZ effect (see discussion below), and Galactic 
residuals, which are reduced by subtracting a template for the Galaxy, $D_\mathrm{T}$. 
The coefficient $C_{\nu}$ is found by minimizing the residuals in the region of the $\mathrm{SZ}_{\nu}$ map not containing 
SZ signal (i.e., known clusters). 
The dust-cleaned map, $M_{217}^\mathrm{c}$, is defined as $M_{217}^\mathrm{c} = M_{217} - A\,(M_{353} - M_{100})$, where the 
coefficient $A$ is found by minimizing the Galaxy-dominated variance of the difference $M_{217}^\mathrm{c} - M_{100}$. 
We find that a value of $A=0.142$ produces a map at 217\,GHz with a minimal amount of residual Galactic foreground in 
the Virgo region (in a different region the optimal value for $A$ may differ).   
We have tested that the value of the coefficient $A$ remains stable when using the CMB from one of the official 
\Planck\ CMB maps \citep[from the {\it SEVEM} method][]{Leach2008,FernandezCobos2012} instead of $M_{100}$. 
However, it is important to note that at the positions of the clusters, the {\it SEVEM} CMB template 
(and the other official \Planck\ CMB templates) are still contaminated by SZ signal.   
An alternative definition of $M_{217}^\mathrm{c}$, using the $M_{143}$ map instead of $M_{100}$, gives 
nearly identical  results.
Since $M_{217}^\mathrm{c}$ involves the difference $M_{353} - M_{100}$ 
(or alternatively $M_{353} - M_{143}$), this difference contains a significant amount of SZ signal that needs to be accounted for in the final SZ map.  
We account for the SZ effect contained in this difference in the final SZ effect map when combining the individual SZ effect maps and transforming into Compton parameter units. 
Finally, the Galaxy template is given by $D_\mathrm{T}=M_{353} - M_{217}^\mathrm{c}$. 
As above, this difference also contains a non-negligible amount of SZ effect that is 
accounted for in the final SZ effect map. 
In Fig.~\ref{Fig_MapDiffs} we show some of these differences, where the dust contribution can be appreciated even 
at low frequencies (see, e.g., the $143 - 100$\,GHz difference). 
Although the Galaxy template removes most of the Galactic component, some Galactic residuals still persist and may 
contaminate the SZ signal. These residuals are discussed below in Sect.~\ref{sect_GalactContam}. 

The Galaxy template differs from more standard ILC methods that would consider 
the high-frequency bands at 545 and 857\,GHz (or a model based on a combination of bands) 
as a tracer of the dust at lower frequencies. 
Our choice is mostly motivated by the variations in the temperature and spectral index of the dust 
across our field of view and also by the fact that low temperature Galactic regions (which are poorly constrained 
if they are too faint) may become more important at lower frequencies.
Instead of using a Galaxy template defined at high frequencies, we use the Galaxy 
template defined by the difference between the 353\,GHz and 217\,GHz bands. 
This difference is the closest in frequency to the key SZ effect frequencies (70, 100, and 143\,GHz). 
When cleaning the 217\,GHz band from dust, we use a different definition for the Galaxy template, taking instead 
the difference $M_{353} - M_{100}$ (or $M_{353} - M_{143}$). 
This is done to avoid introducing correlated noise between the 217\,GHz band 
and the Galaxy template. 
Using the high-frequency bands of \Planck\ at 545 and 857\,GHz as Galaxy templates has the 
advantage of yielding a higher S/N and relatively smaller CMB and SZ effect components, but 
would introduce unwanted Galactic residuals in those Galactic regions with lower temperatures 
(which are not mapped well by the high frequency bands). 
Relying on the difference $M_{353} - M_{217}$ (or $M_{353} - M_{100}$) minimizes this 
problem, since this difference better captures the colder Galactic features present at lower frequencies. 
Given the large area of Virgo and variations in the dust properties, 
it is safer to extrapolate a noisier estimate of the Galaxy from 353\,GHz to 217\,GHz than to extrapolate a 
higher-S/N version of the Galaxy (but also less accurate in colder regions) from 545\,GHz (or 857\,GHz) to 217\,GHz and below. 

\subsection{Final SZ map}
%%%%%%%%%%%%%%%%%%%%%%%%%%
After building the individual SZ effect maps at 70, 100, and 143\,GHZ from Eq.~\eqref{eq_SZ effect}, 
these maps are combined with weights proportional to the inverse of their variance to produce a single SZ effect map. 
The weighted map is converted to Compton parameter units taking into account the \Planck\ instrumental 
response (i.e., the convolution of the SZ spectrum in the non-relativistic case with the bandpass of \Planck) in each band 
and the band differences involved in defining $M_{217}^\mathrm{c}$, $D_\mathrm{T}$, and $\mathrm{SZ}_{\nu}$, as described above. 

The individual SZ effect maps at 70, 100, and 143\,GHz are shown in Fig.~\ref{Fig_SZ_070_100_143}. 
Virgo is clearly seen in all of them as an extended source a little below the centre of the image. Other clusters 
can also be seen in the same field of view, with Coma being the most prominent in the top part of the images. 
Bright radio and infrared sources (including the superluminous M87 at the centre of Virgo) 
and the central part of Coma have been masked, together with small very bright regions 
of the Galaxy. The mask is built after defining a threshold in the SZ map where the brightest features 
(bright clusters like Coma and bright compact Galactic regions which correspond to the regions with the 
highest optical depth in Fig. \ref{Fig_Dust}) are masked out. 
The maps presented in Fig.~\ref{Fig_SZ_070_100_143} have been smoothed to a resolution of 
$1\pdeg5$ (FWHM) in order to better appreciate the extent of the signal around Virgo and reduce the noise. 
The field of view is $31\pdeg9$ across. Some artefacts can be seen in the maps, especially at 143\,GHz, where 
vertical stripes can be clearly appreciated over the entire field of view. We apply a simple technique to reduce 
the negative impact of these stripes. Details of the stripe removal process are given in the Appendix. 
After the stripe removal, the individual SZ maps (at 70, 100 and 143 GHz) are combined together as described above 
to produce the final SZ map (see Figs.~\ref{Fig_SZ_Grid} and ~\ref{Fig_SZ_Contours}).

%%%%%%%%%%%%%%%%%%%%%%%%%%%%%%%%%%%%%%%%%%%%%%%%%%%%%%%%%%%%%
\subsection{Galactic residuals}\label{sect_GalactContam}  
%%%%%%%%%%%%%%%%%%%%%%%%%%%%%%%%%%%%%%%%%%%%%%%%%%%%%%%%%%%%%
%==================================================

%__________________________________________________________________
\begin{figure} % MADE BY: Plot_Ne_Plot_Ne_Ghizzardi04_Urban11_Fit.pro
   \centering
   \includegraphics[width=9.3cm]{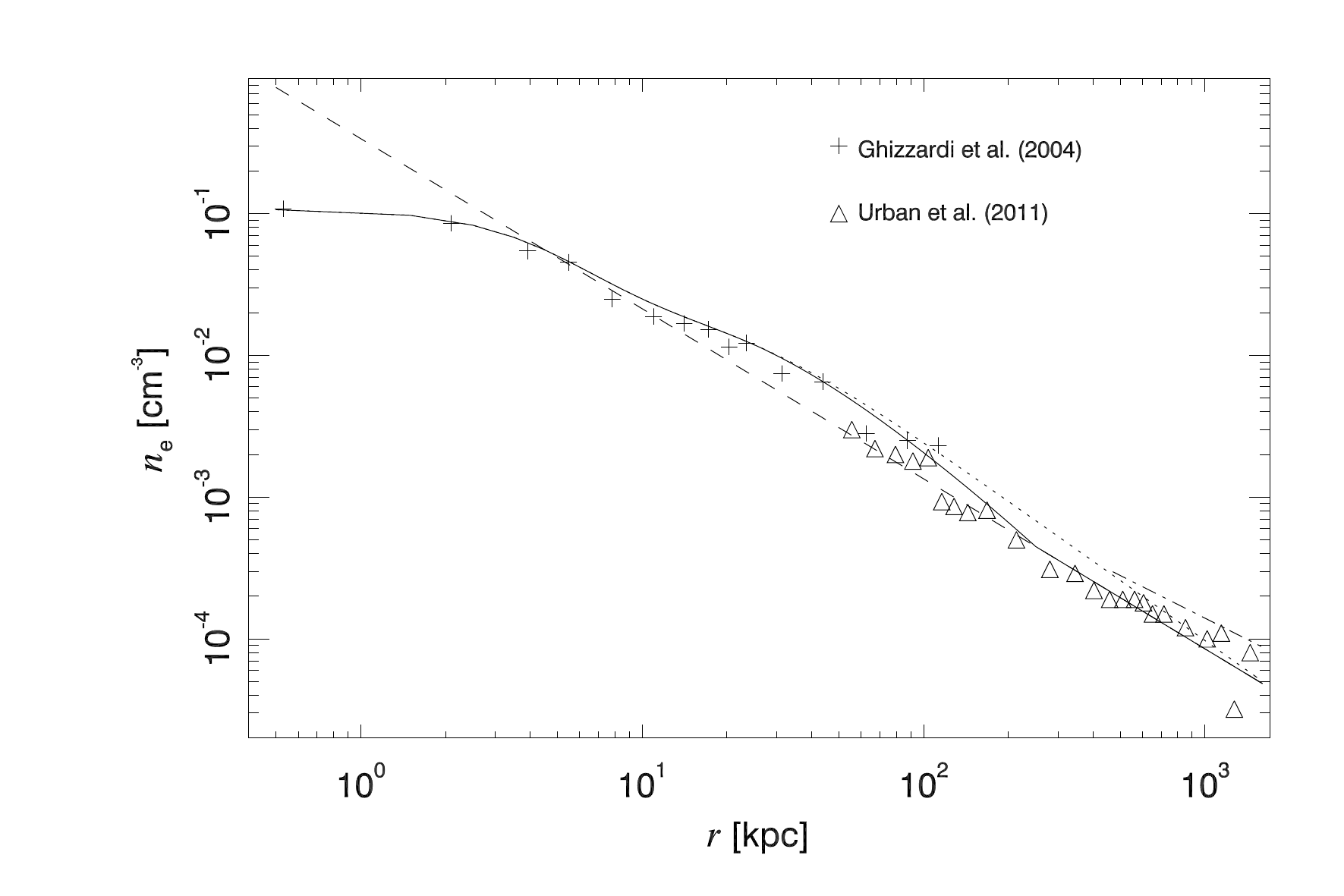}
      \caption{Electron density profile of Virgo derived from \CHANDRA\ and \XMM\ data. 
               The dotted and dashed lines correspond to the fits derived by 
               G04 and U11, while the 
               solid line shows the hybrid model adopted in this paper, combining the features 
               from the G04 and U11 models. The dot-dashed line shows an alternative model 
               that extends the G05 model beyond 500\,kpc with an $r^{-1}$ power law. Error bars are omitted 
               for clarity.}
         \label{Fig_Model_Ne}
   \end{figure}
%__________________________________________________________________

%__________________________________________________________________
\begin{figure} % MADE BY: Plot_Ne_Plot_Ne_Ghizzardi04_Urban11_Fit.pro
   \centering
   \includegraphics[width=9.3cm]{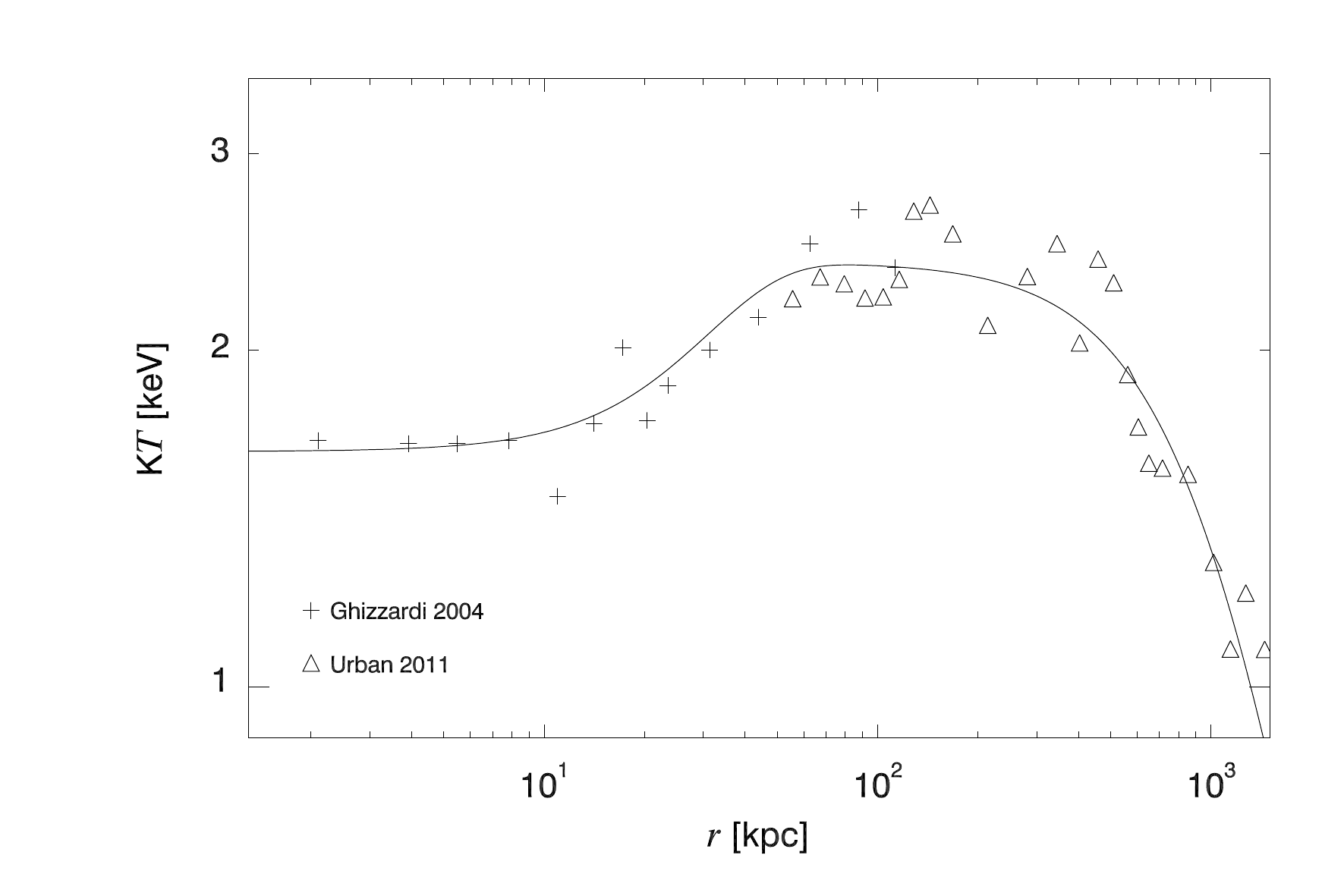}
      \caption{Temperature profile of Virgo derived from \CHANDRA\ and \XMM\ data. 
               The solid line shows the hybrid model adopted in this paper.}
         \label{Fig_Model_T}
   \end{figure}
%__________________________________________________________________

Among the possible systematics affecting our results, the Galaxy is probably the dominant one. Unknown variations in 
the dust temperature and spectral index, especially in diffuse cold regions with low emissivity at infrared frequencies, may 
leave residuals that could be mistaken for, or contaminate, the SZ effect map. 
A simple test can be performed to identify regions where the SZ map is more likely to be contaminated by Galactic 
residuals. We multiply the mean-subtracted maps of the SZ by an estimate of the dust in order to highlight potential Galactic contaminants. The result is shown in Fig.~\ref{Fig_SZ_Dust_correl}. 
We build the dust template as the difference between the 353\,GHz and 217\,GHz bands. 
This choice guarantees that the dust template is as close as 
possible to the dust components used to build the SZ map (70--217\,GHz), thus minimizing 
the bias due to deviations from the mean of the 
dust temperature and spectral index observed near the Virgo cluster (see Fig.~\ref{Fig_Dust}). 
A small amount of SZ effect (compared with the amplitude of the Galaxy) is expected to contribute to this dust 
template but its significance is small for our purposes. We multiply the mean-subtracted SZ map and dust 
template (both at a common resolution of 15\arcm) and later we degrade this map to a resolution of 1\pdeg5. 
The smoothing kernel locally integrates the correlation between the two maps, cancelling out the 
signal in those regions with small correlation and boosting the signal in those regions with more correlation 
(on scales comparable to the smoothing kernel). 
The sequence of dark and bright spots in Fig.~\ref{Fig_SZ_Dust_correl}, which are aligned in the diagonal arc 
(going from the bottom-left to the top-right) identify a region where the SZ map is most likely to be 
contaminated by Galactic residuals. The brightest spot in this arc corresponds to a 6\,$\sigma$ deviation,  
where the Galaxy removal has over-subtracted the Galaxy (mimicking the SZ effect decrement),  while 
the darker regions in the arc mark the regions where the Galaxy removal has not subtraced enough Galactic signal. 
Note how the arc-like diagonal feature seen in Fig.~\ref{Fig_SZ_Dust_correl} traces the cold Galactic regions 
(see also Fig.~\ref{Fig_Dust}).  
The dark blob at the position of Coma (anti-correlation) is produced by the lack of Galactic signal in this region. 

This simple test suggests that the bright 6\,$\sigma$ blob observed in the SZ map in the northern part of Virgo
(and marked with a small circle in Fig.~\ref{Fig_SZ_Contours}) may be a Galactic residual. 
We henceforth mask this area in our analysis to avoid this possible contamination. 
The remaining area of Virgo, on the other hand, seems to be free of significant Galactic residuals. 

%%%%%%%%%%%%%%%%%%%%%%%%%%%%%%%%%%%%%%
\section{Complementing Planck with X-ray data}\label{sect_xrays}
%%%%%%%%%%%%%%%%%%%%%%%%%%%%%%%%%%%%%%
%=====================================

The combination of SZ effect and X-ray data improves the constraints on the plasma properties, as shown in 
\cite{PlanckPressure}. 
We comine several X-ray data sets with \Planck\ to extract additional information about 
the cluster gas distribution. Among other experiments, Virgo has been observed by \ROSAT, \XMM, and \CHANDRA.  
Each of these instruments offers different advantages. 
The \ROSAT\ All Sky Survey (RASS) is to date the most detailed all-sky map of the 
X-ray Universe at low energies (1--2\,keV) and the only experiment that covers the entire cluster. 
In RASS data, Virgo is the largest single extragalactic object in the sky, in terms of its angular size.

\subsection{X-ray data}
%%%%%%%%%%%%%%%%%%%%%%%%%%

We focus on the diffuse \ROSAT\ maps, where the most prominent point sources (AGN above 0.02\,counts s$^{-1}$ in the R6 + R7 band 
or $\approx$ 0.6--1.3 keV, \citealt{Snowden1997}) are masked in the \ROSAT\ maps. Above this threshold, 
there are still some sources that were not completly removed from the data, but their impact in the 
diffuse X-ray maps is small. The diffuse RASS data is presented in a set of maps and bands 
($0.1$--$2$\,keV). 
The low-energy bands are more contaminated by local emission (the Local Bubble and the Milky Way), 
while the high-energy bands contain a more significant contribution from extragalactic active galactic nuclei (AGN) 
that lie below the 0.02\,counts s$^{-1}$ threshold. 
For our purposes we choose the band R6 (0.9--1.3\,keV). 
This band is more advantegeous in terms of instrumental response, background contamination, and cluster versus AGN emission. 
The pixel size of the diffuse RASS maps is 12\arcm\ and the units are counts\,s$^{-1}$\,arcmin$^{-2}$. 
The large pixel size 
(a downgrade from the original pixel size of the PSPC camera) is not an obstacle for our purposes and it is comparable 
to the resolution of our SZ effect map (15\arcm). 

We repixelize the diffuse all-sky maps from \ROSAT\ R6 band (with a native pixel size of $12\arcm\times12\arcm$) 
%using {\small HEALPIX} at the resolution level Nside = 2048.
using {\tt HEALPix}\footnote{\citet{gorski2005}, \url{http://healpix.sf.net}} 
at the resolution level $N_\mathrm{side} = 2048$.
Although the R6 band is the best compromise between Galactic and AGN contamination, 
it still contains Galactic emission that affects the Virgo region. 
In particular, the North Galactic Spur reaches the vicinity 
of the Virgo cluster. We mask this region from our analysis of the X-ray data (when computing the profiles).  

In addition to \ROSAT\, we use results from two other observatories. 
\XMM\ data have been used recently by U11 to study the X-ray emission in Virgo out to the virial radius. 
Although the \XMM\ data used in 
that work covers only a linear slice through a small fraction of the cluster area, it is of great interest 
since it constrains the gas density and temperature out to large radii. 
Figure~\ref{Fig_Planck_vs_RASS} 
shows the sector used in this \XMM\ study superimposed on the SZ effect and RASS maps. 

At higher-resolution, \CHANDRA\ data have been used by G04 to study in detail the innermost 
region of Virgo, producing tight constraints on both 
electron density and temperature. 
Figure~\ref{Fig_Model_Ne} shows these two data sets compared with models for the electron density. 
In addition, we also show a model that is built from the combination of the two models (solid line), which we 
refer to as the ``hybrid'' model. 
The temperature profile is shown in Fig.~\ref{Fig_Model_T}; the solid line shows the hybrid model for the temperature, which we discuss in Sect.~\ref{Model_Results}. 

\subsection{X-ray to $\mathrm{SZ}$ ratio}
%%%%%%%%%%%%%%%%%%%%%%%%%%%%%%%

%__________________________________________________________________
\begin{figure}
   \centering
   \includegraphics[width=9cm]{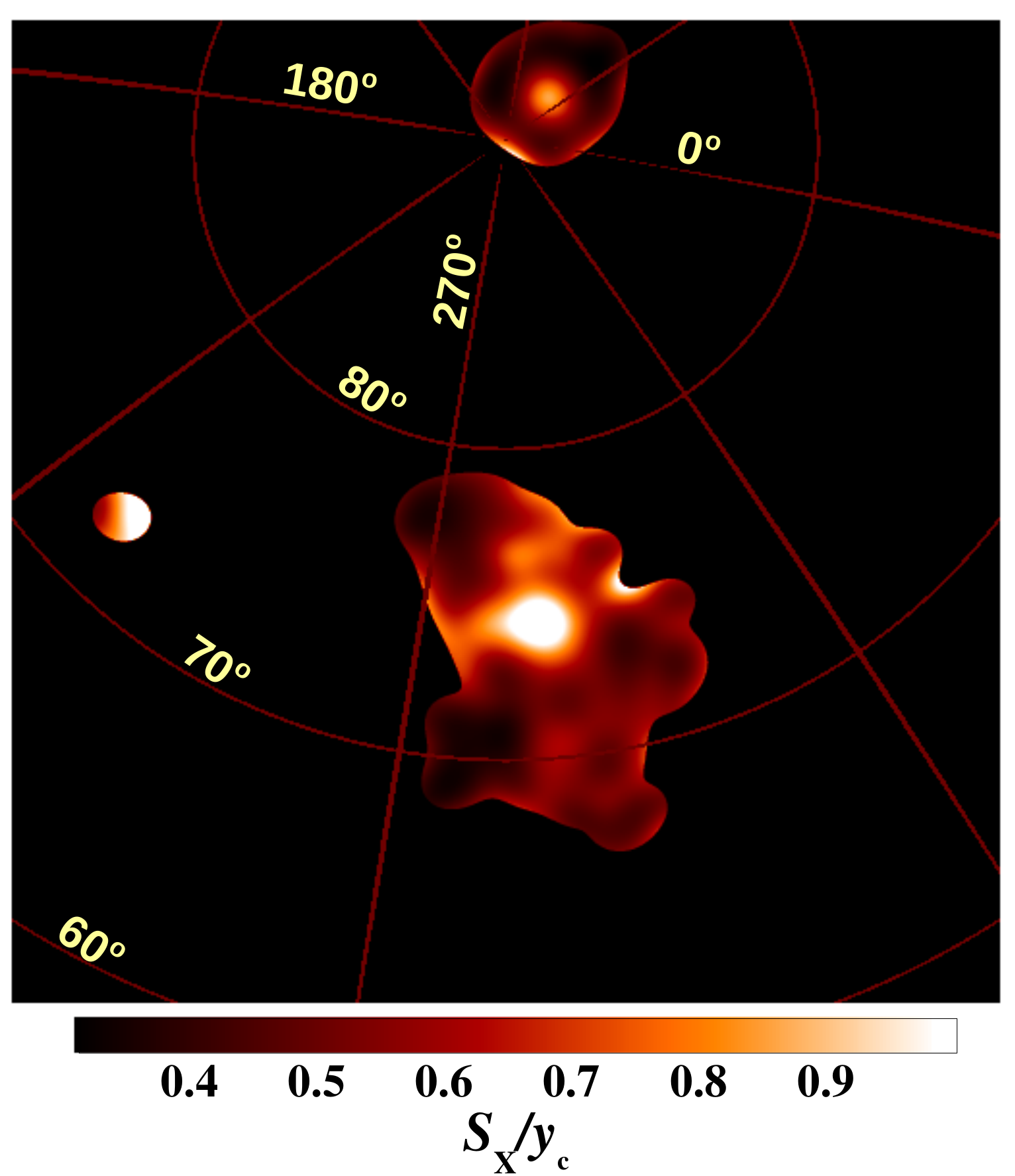} % Made by Virgo_Planck/Plot_XR2SZ_ratio.pro (balder)
      \caption{Ratio between \ROSAT\ and \Planck\ data (smoothed to $1\pdeg5$). The ratio $S_{\rm X}/y_{\rm c}$ is defined only in regions 
               with sufficient SZ signal (the smoothed signal is larger than 1.5 times the dispersion of the map). 
               Three clusters can be distinguished in this plot, Coma in the north, Virgo in the 
               centre, and Leo in the east. The ratio is normalized to 1 at the maximum.}
         \label{Fig_XR2SZ_ratio}
\end{figure}
%__________________________________________________________________

In Fig.~\ref{Fig_XR2SZ_ratio} we present the 
the ratio of the X-ray flux,  $S_{\rm X}$, to the Compton parameter map of Virgo, $y_{\rm c}$, (the Coma and Leo clusters are also seen in the map).
It shows the ratio of smoothed versions (Gaussian kernel with $\mathrm{FWHM}=1.5\deg$) of the X-ray and the $y$-maps
for regions where the smoothed $y$-map S/N is above 1.5. The increase of the ratio at the edge of the 
contour region is mostly driven by the decrease in the $y_{\rm c}$ signal. 
This map shows crudely (but free from any geometrical assumptions) the electron density profile
of the clusters in the plane of the sky. This can be seen from 
\begin{align}
\frac{S_{\mathrm{X}}}{y_{\rm c}}{(\theta)} & \propto \frac{\int dl\, n_{\rm e}^2{(\theta)}\,\Lambda{(k_\mathrm{B}T{(\theta)})}}{\int dl\, n_{\rm e}{(\theta)}\, k_\mathrm{B}T{(\theta)}}\nonumber\\
 & = \left.\frac{\int dy_{\rm c}{(\theta)}\,(n_{\rm e}\,\Lambda{(k_\mathrm{B}T)/k_\mathrm{B}T)}{(\theta)}}{\int dy_{\rm c}{(\theta)}}\right|_{dy_{\rm c}{(\theta)}\propto n_{\rm e}{(\theta)}\, k_\mathrm{B}T{(\theta)}\, dz}\nonumber\\
 & = \langle n_{\rm e}\,\Lambda{(k_\mathrm{B}T)}/(k_\mathrm{B}T)\rangle_{y_{\rm c}}(\theta),
\end{align}
where $\Lambda{(k_\mathrm{B}T)}$ is the temperature dependence of the X-ray emission
in the observed band and $\theta$ is the direction of the line of sight. 
For an isothermal cluster, this map shows the
$y_{\rm c}$-weighted (or equivalently pressure-weighted) electron density
along each line of sight. Since the temperature within the cluster gas
varies at most by a factor of a few, it only weakly modulates the X/$y_{\rm c}$-map,
in comparison to the electron density, which exhibits a much steeper variation.
Furthermore, since the cluster pressure profile is highly structured
and is typically larger in the plane of the sky going through the cluster
center, then the X/$y_{\rm c}$-map predominantly shows the electron density within
this plane. 
Although this interpretation is certainly complicated by temperature
structures in the cluster gas (or errors in the X-ray and more importantly in the SZ map), 
as well as by the remaining impact of the pressure weighting, the X/$y_{\rm c}$-map provides a simple and robust
way to obtain some insights into the structure of the cluster plasma.
%In Fig XXXX we show the (rescaled) spherically averaged profile of
%the X/y-map in comparison with the fitted and de-projected electron
%density profiles of Virgo. 

%%%%%%%%%%%%%%%%%%%%%%%%%%%%%%%%%%%%%%
\section{Electron density and temperature Models}\label{sect_models}
%%%%%%%%%%%%%%%%%%%%%%%%%%%%%%%%%%%%%%
%=====================================
To model the electron density, we use two models found in the literature 
and derived from \CHANDRA\ and \XMM\ data. 
The first one is a model derived by G04 (equations 2 and 3 in that paper), which  
describes the data in the central $r<120$\,kpc region. This model does not account for the cluster emission in the outer regions of Virgo, but 
should be an accurate description of the central part. For the outer regions we consider a second approach that corresponds to the sector model derived by U11, which extends the profile up to the virial 
radius ($\approx 1$ Mpc). 

In addition to these two models, in some later parts of this paper we use a hybrid model that combines the features 
from G04 and U11. 
Following U11, we model the electron density in the outer region as a power law,
\begin{equation}
n_{\mathrm{e}}(r) = \frac{8.5\times10^{-5}}{(r/\mathrm{Mpc})^{1.2}}\,{\rm cm}^{-3}.
\label{eq_ne}
\end{equation} 
This power law is shown as a dashed line in Fig.~\ref{Fig_Model_Ne}.
The inner region is described by the model of G04, while the intermediate (approximately 30--250 kpc) 
region between the G04 and U11 estimates is linearly interpolated between the G04 and U11 models. This hybrid model 
is shown as a solid line in Fig.~\ref{Fig_Model_Ne}. 
Alternatively, in Sect.~\ref{Beyond}, we use an additional model that assumes the density profile from G04 
up to $r=500$\,kpc and beyond this radius the density goes like an $r^{-1}$ power law. This is shown as the  
dot-dashed line in Fig.~\ref{Fig_Model_Ne}.

%__________________________________________________________________
\begin{figure} % MADE BY: Eugene
   \centering
   \includegraphics[width=9.3cm]{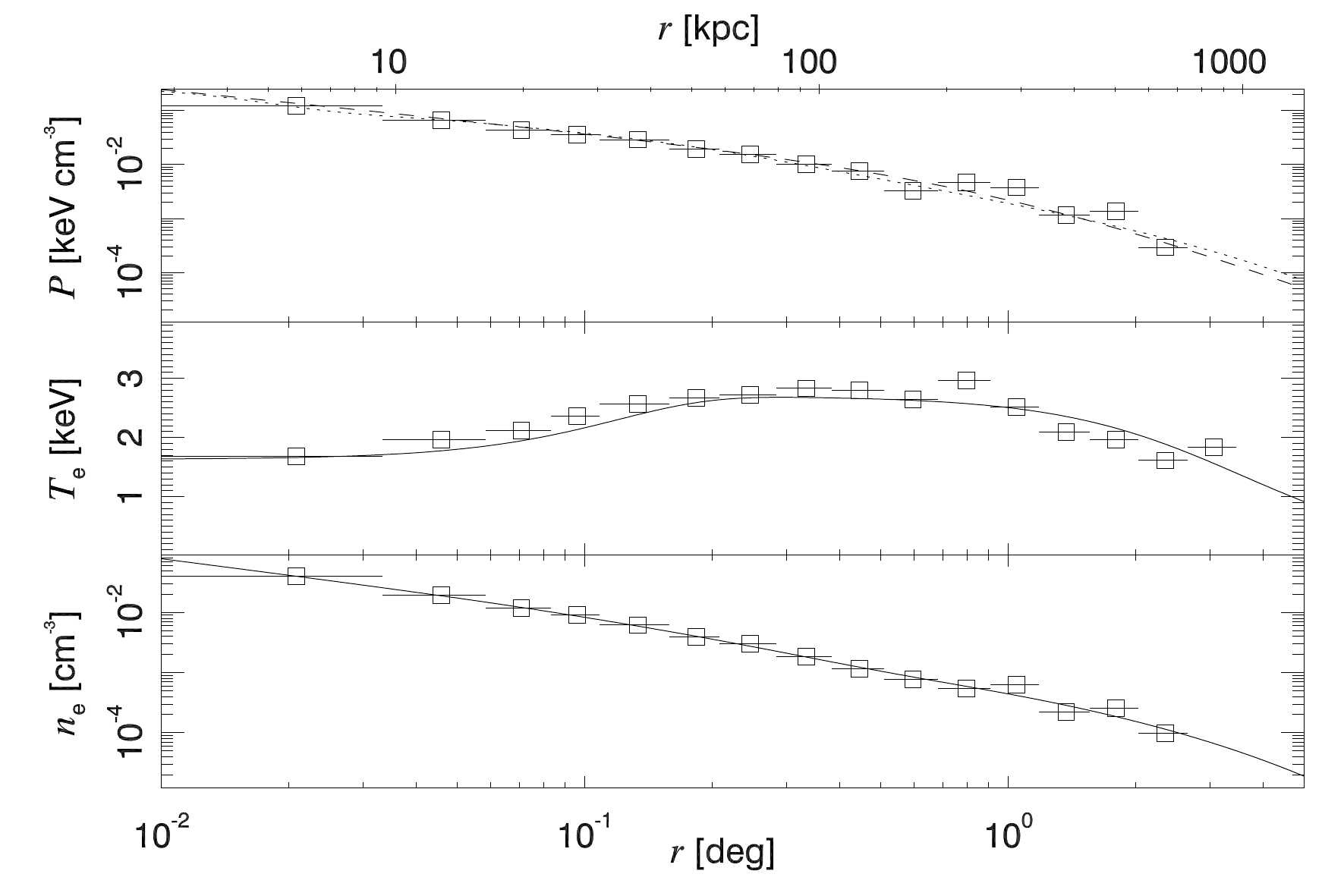}
      \caption{Deprojected pressure (top) and electron density profiles (bottom) 
               from the sparse XMM data. %(see insert in Fig.~\ref{Fig_PressureProfile}). 
               The middle panel shows the projected temperature (see text).  
               The pressure (top panel) is compared with an Arnaud et al. (2010) profile (dashed line), which fits the 
               data well within the virial radius. The solid curves (dotted curve in the case of the pressure) 
               represent a simple analytical model that fits the derived 
               temperature and electrion density. The pressure predicted from the deprojection analysis (dotted curve)
               is compared with the observations in Fig.~\ref{Fig_PressureProfile}. }
         \label{Fig_deproj}
   \end{figure}
%__________________________________________________________________

%__________________________________________________________________
\begin{figure} % MADE BY: Eugene
   \centering
   \includegraphics[width=9.3cm]{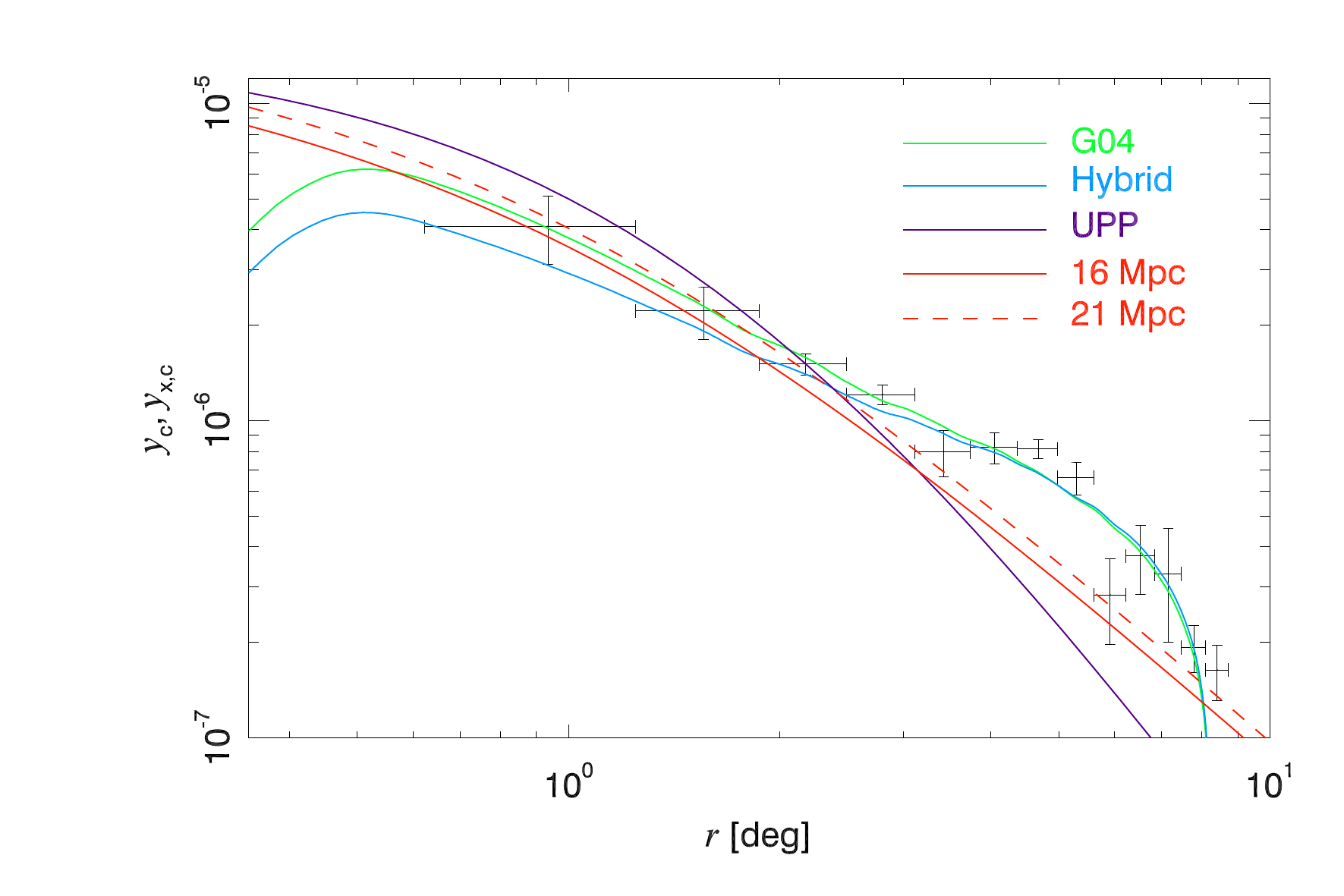}
      \caption{Binned pressure profiles derived for Virgo (black points with errors). 
               The light-blue line shows the prediction from the hybrid model and the green line 
               shows the prediction from the G04 model. Both models decrease at small radii due to the 
               point source mask in \Planck\ data, which is also applied to the simulated data (models). 
               The dashed and solid red curves show the deprojected pressure from X-ray data 
               assuming a universal shallow profile (with outer slope $\beta=3$, see text) for a Virgo distance of
               21\,Mpc and 16\,Mpc, respectively. As a comparison we show as a solid purple line 
               the standard (steeper) \cite{Arnaud2010} profile ($\beta=4.13$) for the case where the distance is 
               set to 16\,Mpc.} 
               %The sparse sampling by \XMM\ used to infer the deprojected pressure is 
               %shown in the bottom left with the circle marking the virial radius. The stripe 
               %with \XMM\ pointings used 
               %in U11 is seen clearly in the vertical direction.}
         \label{Fig_PressureProfile}
   \end{figure}
%__________________________________________________________________

For the temperature profile, we use a model that combines the features from G04 and U11. This model 
is shown as a solid line in Fig.~\ref{Fig_Model_T} and is given by
\begin{equation}
T(r) =  \frac{2.4 - 0.77\times \mathrm{e}^{-(r/\mathrm{Mpc})^2/(2\sigma_r^2)} }
                         {1 + [0.9\times (r/\mathrm{Mpc})]^2 }{\rm keV}\,,
\label{eq_T}
\end{equation}
with $\sigma_r = 23.7$\,kpc. 
%The radius in Eqs.~\eqref{eq_ne} and \eqref{eq_T} is given in Mpc
We should note however that due to the masking of the central M87 region 
of Virgo, 
the results from the hybrid model are very similar to the results derived using the U11 model.

%__________________________________________________________________
\begin{figure*} % MADE BY:MakeProfilesPlot_Planck_ROSAT_Ghizaardi_HybridUrban.pro
   \centering
   \includegraphics[width=18.0cm]{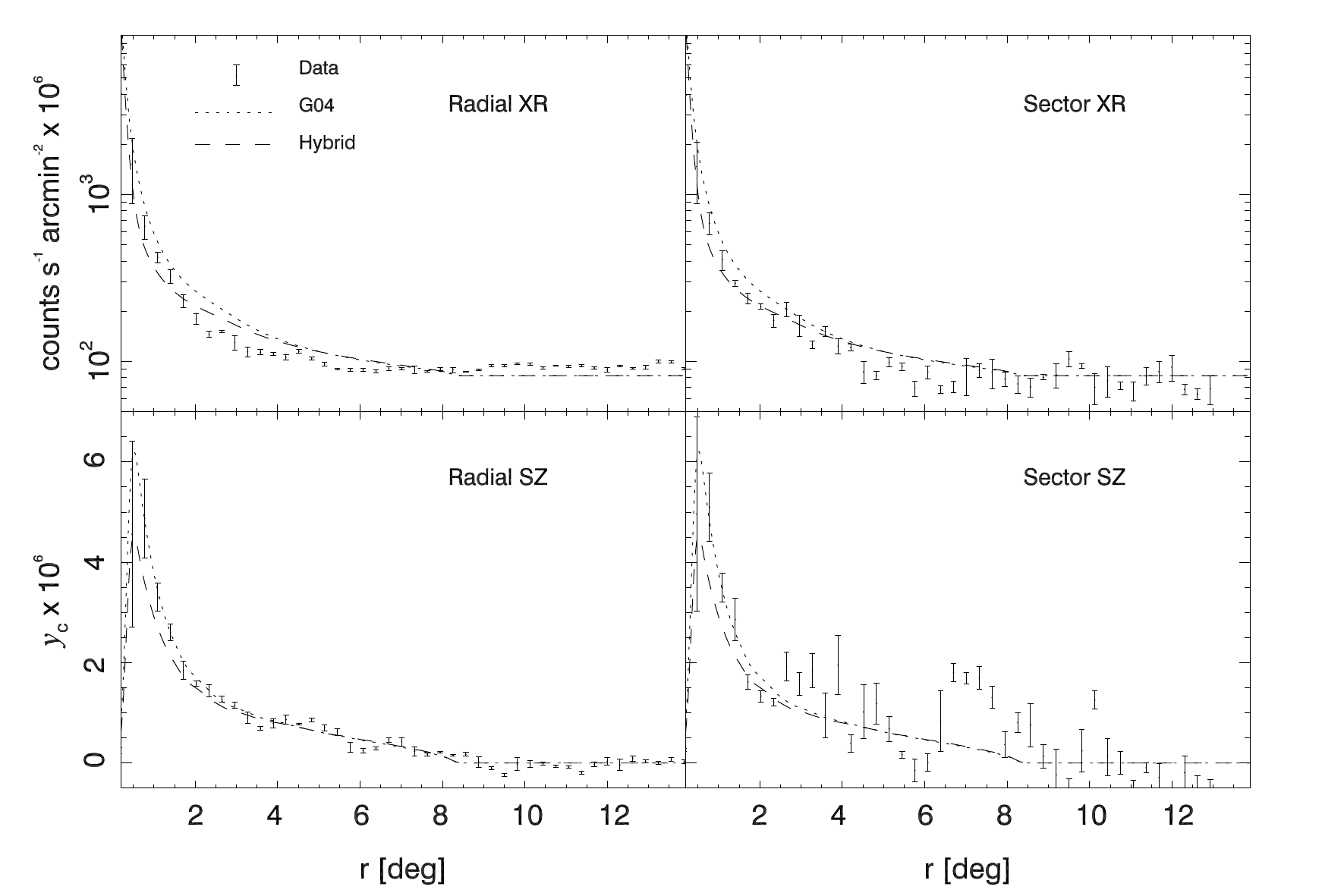}
      \caption{SZ and X-ray profiles compared with the G04 and hybrid models. 
               The two plots on the left correspond to the radial profile and the 
               two plots on the right correspond to the sector profile. For the sector profile, 
               the SZ feature at $7\deg$ is not masked in order to highlight its effect. }
         \label{Fig_Data_vs_Model}
   \end{figure*}
%__________________________________________________________________

At large radii, the model shown in Eq.~\eqref{eq_T} drops quickly. To avoid unphysically low 
temperatures we set a lower limit for the temperature at 1\,keV in the outer region ($r > 1$\,Mpc), 
which is consistent with what is expected between  $R_{\rm vir}$ and $2\times R_{\rm vir}$, 
where the temperature is expected to change only by a small amount \citep{Roncarelli2006,Burns2010}.   
When modelling the cluster, we set a truncation radius at $\approx 8\pdeg3$. 

\subsection{Deprojected models}\label{Sect_deproj}
%%%%%%%%%%%%%%%%%%%%%%%%%%%%%%%%

In addition to these models, and as an independent comparison, we derive our own pressure profile 
 based on a sparse set of \XMM\ pointings that irregularly sample a region of a few degrees around Virgo.  %(see insert in Fig.~\ref{Fig_PressureProfile}). 

For the analysis of the \XMM\ data 
we follow the procedure described by \cite{Churazov2003}, which we briefly summarize here. 
The data are cleaned from flares using the count-rate of the combined MOS detectors above 10\,keV. 
The steady component of the background is calculated from the data of blank field observations provided 
by the XMM SOC site\footnote{\url{http://xmm.esac.esa.int/}}. 
The remaining variable background component is estimated using the mean flare 
spectrum shape (averaged over a large set of moderate and faint flares) and normalized using the 
observed spectrum between 11 and 12\,keV. 
The sum of the steady and variable components in each observation is used as a total background 
in this observation. The image is built in the 0.5--2\,keV band, background-subtracted and vignetting-corrected. For each observation a vignetting map is evaluated for the mean energy of the image, taking 
into account the offset of a given pixel from the pointing centre of the observation. 
The vignetting maps are multiplied by the corresponding exposure times and combined into a {\it vignetted}  
exposure mosaic. For the spectra we follow a slightly different route for the vignetting correction. 
Namely, for each event from the event-list we calculate the vignetting based on the event energy and offset 
from the pointing centre of a given observation. When the spectra are accumulated over a given region, 
the events corrected for this drop in efficiency are added together. 
Appropriate corrections are also applied to the background events and 
to errors associated with each spectral channel.

After the data are corrected, we use the sparse pointings to deproject the emissivity, assuming 
spherical symmetry (see \citealt{Churazov2003}). 
Prior to the deprojection all strong compact sources and few prominent Virgo galaxies 
are removed.

\subsection{Derived profiles}
%%%%%%%%%%%%%%%%%%%%%%%%%%%%%%
In Fig.~\ref{Fig_deproj} we show the electron density, gas temperature, and
pressure radial profiles. Notice that in the middle panel we show the
projected values of the temperature (to avoid large scatter in the
deprojected $T$ values). The density and temperature radial profiles
have been approximated with simple analytic functions, shown with black
solid lines. The top panel of Fig.~\ref{Fig_deproj} shows the product
of the analytic approximations for the density and temperature distributions
(black curve). Also shown is a universal pressure profile
\citep[][blue dashed line]{Arnaud2010} with $P_0=3.5$\,keV\,cm$^{-3}$, $c_{500} = 2.5$,
$\gamma = 0.8$, $\alpha = 1.33$, and with the outer slope fixed to a
shallow index, $\beta = 3.0$, using the form
\begin{equation}
P(x)=\frac{P_0}{(c_{500}\,x)^{\gamma}[1+(c_{500}\,x)^{\alpha}]^{(\beta-\gamma)/\alpha}},   
\end{equation}
with $x=r/r_{500}$. For this profile, we adopt the value $r_{500} = 770$\,kpc (with $r_{500}$ the 
radius at which the enclosed overdensity is 500 times the critical density),  
derived from the X-ray derived temperaure and cluster mass relation, $T_{\rm X}$--$M$, 
from \cite{Vikhlinin2006} for $T_{\rm X} \approx 2.3$\,keV.

We should emphasize again that the deprojection technique involves the
assumption (probably erroneous in this case) that the cluster has
spherical symmetry. The possible geometry of the cluster is discussed in
Sect.~\ref{ProfileSectors}. Also, the sparse coverage of the Virgo
region by \XMM\ is neither random nor systematic (except for the stripe
to the north), but tends to concentrate on cluster member galaxies or
prominent AGNs. These effects leave considerable uncertainty in the
X-ray-based profiles.

%______________________       MODEL  FIT    ____________________________________

%%%%%%%%%%%%%%%%%%%%%%%%%%%%%%%%%%%%%%%%%%%%%%%%%%%%%%%
\section{Model fits to the data}\label{Model_Results}
%%%%%%%%%%%%%%%%%%%%%%%%%%%%%%%%%%%%%%%%%%%%%%%%%%%%%%%
%________________________________________________________________________________

It is interesting to compare the predicted signal from the X-ray emision observed by \XMM\ up to the virial 
radius (see U11). We compare the data with the models by computing a ``radial profile'' (average signal in circular bins) and a ``sector profile'' (average signal along the sector shown in Fig.~\ref{Fig_Planck_vs_RASS}) for both the data and the models. 
The radial profile of the data is computed after smoothing both \Planck\ and \ROSAT\ data with a Gaussian, the final resolution after the smoothing being 15\arcm\ for both data sets. 
For the sector profile, we compute the profiles considering only the data contained in the 
rectangular region of Fig.~\ref{Fig_Planck_vs_RASS} (but extending the rectangle towards the north-east). 
The extension of this region towards the north-east intersects the feature that is likely due to Galactic 
contamination in the SZ map (marked with a small circle in Fig.~\ref{Fig_SZ_Contours}). 
This feature is clearly visible as a bump in the SZ sector profile at 
around $7\deg$. In the X-ray profile from \ROSAT, which extends beyond the region studied by U11, there 
is no evidence for a similar feature, supporting the hypothesis that this feature is a cold Galactic residual 
in the SZ map.  
%use the \ROSAT\  data at the same 15' resolution but 
%we smooth the \Planck\   data to a $1.5^{\circ}$ resolution in order to maximize the 
%signal-to-noise ratio. This sector offers a relatively small area over which to average 
%the \Planck\   data.  

The Compton and X-ray profiles for the G04 and hybrid models (for $n_{\rm e}$ and $T$), are computed assuming spherical symmetry and are truncated at $10\deg$ (or 3.7\,Mpc).  
Virgo is assumed to be at a distance of 21.46\,Mpc ($z=0.005$) and we adopt 
a {\tt MeKal} model with metallicity $Z=0.3$ and HI column density $n_\mathrm{H}= 0.0207\times 10^{22}$\,cm$^{-2}$. This model is convolved with the instrumental response of ROSAT using {\tt XSPEC} at the desired energy range to produce the predicted count rate. We build projected 2D maps (SZ and X-ray) that are masked and smoothed with the same masks and smoothing kernels applied to the real data and described above. 

Figure~\ref{Fig_Data_vs_Model} shows the comparison between the data and the models. 
Errors in the profiles are discussed in more detail in Sect.~\ref{Beyond}. 
The X-ray radial profile is reproduced reasonably well by both models, with the 
model of G04 predicting more signal than the hybrid model. 
%Both models overpredict the observed $X$-ray signal at radii $R>2^{\circ}$. This mismatch 
%could be due to the lack of symmetry of the cluster and the fact that the model at large 
%radii is derived from a single sector in the cluster that may not be representative 
%of the circularly averaged signal at large radii. 
For the sector profile, the hybrid model fits the data better than the G04 model in the intermedite region ($r \approx 2\deg \approx$ 600 kpc). 
The G04 model beyond 200\,kpc (corresponding to about 0\pdeg6) 
has a higher electron density than the U11 model (see Fig.~\ref{Fig_Model_Ne}), which results in more predicted X-rays than 
observed in the sector region at around $2\deg$. 
When comparing the models with the SZ effect data, the differences are more evident. 
In both the radial and sector profiles (bottom panels in Fig.~\ref{Fig_Data_vs_Model}),  
the hybrid model does not describe the central region as well as the G04 model, which gives an excellent 
fit to the core region in both cases. 
In the outer regions both models converge and perform similarly in the radial  
and sector profiles.

%The feature (bump) in the SZ sector profile at $\sim$ $7^{\circ}$ 
%(see bottom right panel in Fig. \ref{Fig_Data_vs_Model}) is due to the (likely) residual Galactic emission 
%in the SZ effect map as discussed in Section \ref{sect_GalactContam} which unfortunately happens 
%to be aligned with the direction of the \XMM\  sector from U11. 

Better S/N in the SZ data can be obtained by binning the SZ signal in wider bins. 
Figure~\ref{Fig_PressureProfile} shows the Compton parameter, $y_{\rm c}$, 
from \Planck\ in wide radial bins (33\arcm\ wide). The errors are computed 
as the disperion of the individual radial bins in each wide bin (see Sect.~\ref{Rad_Bins} for more details).
Also shown is the signal predicted by the shallow models described in Sect.~\ref{sect_models}, namely the G04 model (green line) and the hybrid model (light blue line). It is important to emphasize that these models are 
generally constrained with data up to around $4\deg$, so the models beyond this point are extrapolations. 
In addition, the shallow $y_{\rm c}$ profiles ($\beta=3$) inferred from the deprojection analysis (from
the sparse X-ray data) are shown as red curves. The solid one assumes that M87 is at a distance of
$D=16$\,Mpc, while the dotted line assumes $D=21$\,Mpc. For comparison the
purple solid line corresponds to the \cite{Arnaud2010} profile with the standard set
of parameters; $P_0=6.41$ keV/cm$^3$, $c_{500} = 1.81$, $\gamma = 0.31$, 
$\alpha =1.33$, and $\beta = 4.13$. Both versions of the \cite{Arnaud2010} profiles 
(shallow index $\beta = 3.0$ and standard values) 
give a reasonable description to the observed profile in the central region 
(up to around $2.5\deg$, corresponding to $r_{500}$)
but underpredict the observed (SZ) pressure at the virial radius
(roughly $4\deg$) and beyond, with the shallow profile ($\beta = 3.0$) 
reproducing the data better.

%__________________________________________________________________
\begin{figure} % MADE BY: MakeVirgoProfile_BinnedErrors.pro
   \centering
   \includegraphics[width=9.3cm]{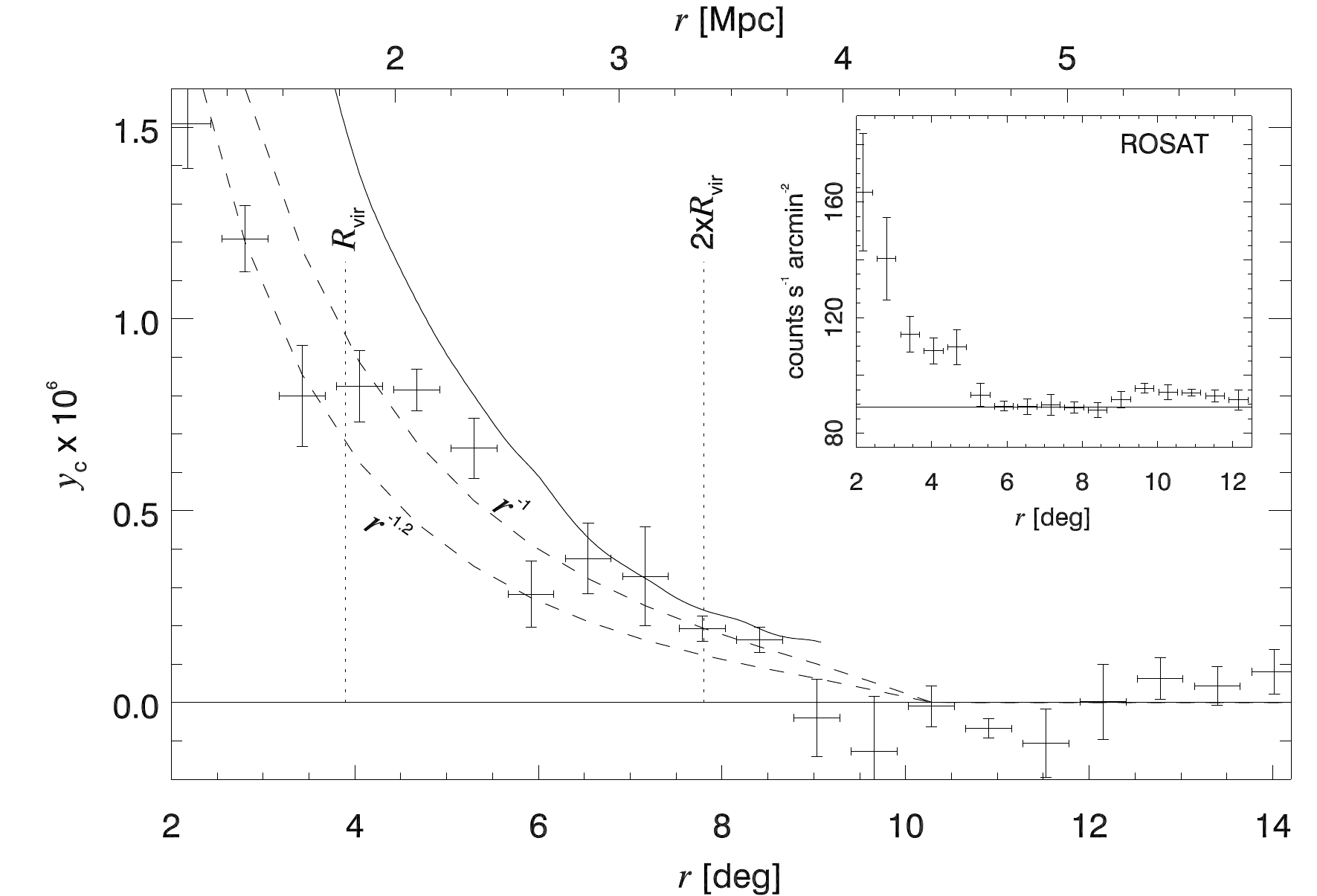}
   \caption{Binned radial $y_{\rm c}$ profile of Virgo. 
               The bins are 30 pixels wide (or equivalently 33\arcm\ , or 230\,kpc). The two 
               dashed lines correspond to the expected signal from two analytical models where the electron 
               density beyond 0.5\,Mpc falls as $r^{-1}$ or $r^{-1.2}$. 
               At $r=0.5$\,Mpc the model is normalized 
               to the electron density given by G04 at that distance. At distances smaller than 0.5\,Mpc 
               the electron density follows the model of G04. The solid line shows the profile of the simulated Virgo 
               cluster. 
               The inset shows the corresponding profile in 
               the \ROSAT\ map using the same bins (with units the same as in Fig.~\ref{Fig_Data_vs_Model})}
         \label{Fig_BinnedProfile}
   \end{figure}
%__________________________________________________________________

An indication that the \cite{Arnaud2010} profile (in its standard form) derived from X-ray data 
underpredicts the pressure at large radii was also found in \cite{PlanckPressure}, where a joint 
X-ray and SZ analysis concluded that 
shallower profiles ($\beta < 4.5$ and $\beta < 4$ for non cool-core clusters) are favoured. 
A similar conclusion was obtained in the \Planck\ analysis of the Coma cluster \citep{PlanckComa}. 
In the case of Coma, it was found that even shallower pressure profiles 
($\beta \approx 3$) were required to fit the SZ signal at large radii \citep{PlanckComa}. The Virgo cluster 
allows us to extend the radii even further and points in the direction of an increased shallowness at the largest radii. 
This is expected at some point, since the $y_{\rm c}$ profile should converge to the average $y_{\rm c}$ for 
sufficiently large radii. If reservoirs of ionized gas concentrate around the clusters (as shown in Fig.~\ref{Fig_Sim_Virgo}) we should expect a gradual flattening of the profile as the distance to the cluster centre 
increases.
 
Another interesting result is the fact that the G04 model seems to agree better in the central region, while both 
models agree well with the observations in the most distant regions. Both models go to zero at the truncation radius 
(about $8\pdeg3$). The SZ signal extends beyond twice the virial radius suggesting the presence of warm gas 
around the cluster. The signal in the outer regions is studied in more detail in Sect.~\ref {Beyond}. 
Finally, the two profiles derived from the sparse \XMM\ pointings show some dependence on the assumed distance 
to Virgo. This will be discussed in more detail 
in Sect.~\ref{subsect_dist}.

%%%%%%%%%%%%%%%%%%%%%%%%%%%%%%%%%%%%%%%%%%%%%%%%%%%%%%%%
\section{Gas beyond the virial radius}\label{Beyond}
%%%%%%%%%%%%%%%%%%%%%%%%%%%%%%%%%%%%%%%%%%%%%%%%%%%%%%%%
%________________________________________________________________________________

Figure~\ref{Fig_SZ_Contours} shows how the SZ signal extends beyond the virial 
radius. This picture is quite different for X-ray observations, where the emission vanishes soon after 
the virial radius. The lack of X-ray signal in regions where there is still 
SZ signal can be used to constrain the properties of the plasma in these regions. Intuitively, 
a boost in the SZ signal from a low-density plasma can be produced (without a 
significant X-ray counterpart) if the plasma  extends along the line of sight 
as a low-density medium and/or its temperature remains sufficiently high (1\,keV or above). 
In the previous section, a preliminary comparison between the SZ and X-ray profiles was presented. 
In this section we focus on the signal at the outskirts of the cluster, where the gas properties 
are poorly constrained.

\begin{figure} % MADE BY: PlotVirgoSZDX11_Contours.pro
   \centering
   \includegraphics[width=9.cm]{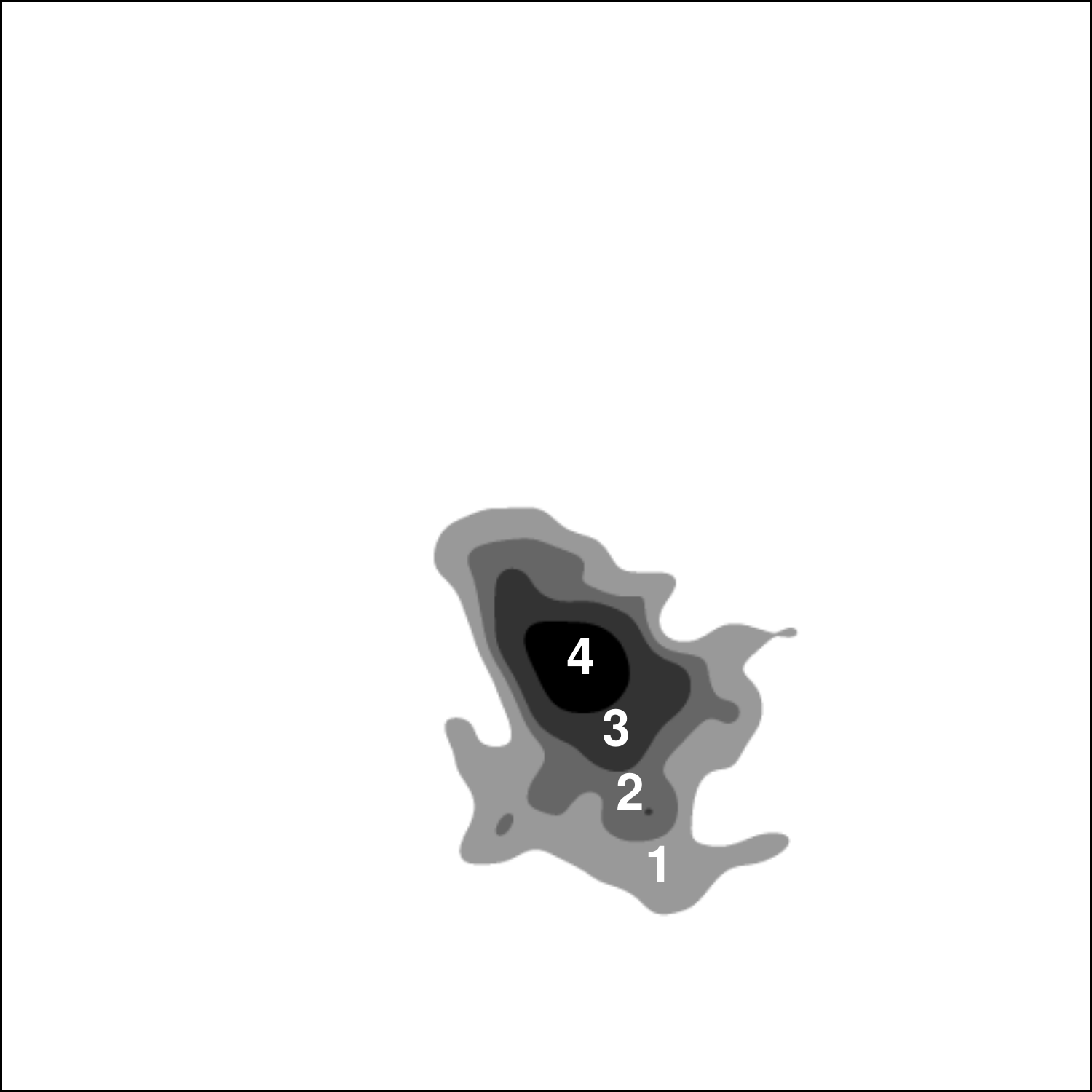}
      \caption{Four regions defined by the S/N of the SZ effect. 
               From lightest to darkest, $\mathrm{S/N} \in [2,3)$,  $\mathrm{S/N} \in [3,4)$, $\mathrm{S/N} \in [4,7)$, and 
               $\mathrm{S/N} > 7$. Note that the likely spurious structure in the northern part of Virgo is not 
               included in our analysis. See Fig.~\ref{Fig_SZ_Contours} to see the same contours 
               in relation to the SZ signal.}
         \label{Fig_Regions}
   \end{figure}

We study the signal in the outskirts of Virgo in two different ways. First, we stack 
the signal in wide radial circular bins around the central point. We refer to these bins as ``wide-radial bins''. 
Second, we account for the lack of symmetry in Virgo and define non-symmetric 
bins that trace the S/N of the SZ map (see Fig.~\ref{Fig_Regions}). We refer to these bins as ``contour-region bins''.
 
\subsection{Wide-radial bins}\label{Rad_Bins}
%########################
In order to maximize the S/N in a radial bin we consider wide bins with a 33\arcm\ gap 
between the inner and outer radius of each bin (or 30 pixels wide in our native maps). In each wide bin we compute 
the mean and dispersion of the individual profiles (the individual profiles are computed in a ring of width 1 pixel).  
These profiles are computed considering only regions in the SZ map that are not masked by the point source 
or Galactic mask. The Galactic mask also excludes from our analysis the region 
in the north part of Virgo 
(small circle in Fig.~\ref{Fig_SZ_Contours} or 6\,$\sigma$ fluctuation in Fig.~\ref{Fig_SZ_Dust_correl}) 
that is likely contaminated by Galactic dust (see Sect.~\ref{sect_GalactContam} above). 
The resulting binned profile, with its corresponding dispersion, is shown in Fig.~\ref{Fig_BinnedProfile}. We should emphasize that the error bars shown in Fig.~\ref{Fig_BinnedProfile} represent the dispersion of the individual 1 pixel wide radial bins in the 30 pixel wide ring. This dispersion accounts for the correlated nature of the noise 
and the intrinsic variability of the profile in the 33\arcm\ angular range.  
The binned profile shows an excess with respect to the background out to twice the virial radius 
(and possibly beyond). 

%__________________________________________________________________
%__________________________________________________________________

For comparison, we also show in the inset the profile from \ROSAT\ using the same radial bins. A first  
conclusion is that the SZ signal seems to extend significantly farther than could be inferred from X-ray data alone, 
directly pointing to the presence of previously undetected plasma at large radii. 
Also, interestingly, the average profile is consistent with the extrapolation of the U11 profile ($\propto r^{-1.2}$). 
For comparison, we also show as a dashed line in Fig.~\ref{Fig_BinnedProfile} the expected signal 
for a shallower ($r^{-1.0}$) profile. 
This model assumes, for the electron density, 
the G04 model up to $r=500$\,kpc and an $r^{-1.0}$ profile beyond this point.
The $r^{-1.0}$ power law part of this model is shown in Fig.~\ref{Fig_Model_Ne} as a dot-dashed line.
%However, we should note that the $r^{-1.2}$ profile considered in Fig.  
%\ref{Fig_BinnedProfile} already overpredicts the average $X$-ray signal in the outskirts of Virgo as shown in Fig.  
%\ref{Fig_Data_vs_Model} so the $r^{-1.0}$ would be in even more tension with the $X$-rays. 
The shallow SZ profile beyond the virial radius is significantly different from the average profile observed 
in the outer radii of other \Planck\ clusters \citep{PlanckPressure}, where steeper profiles closer to $r^{-3}$ or even $r^{-4.5}$ 
are preferred beyond the virial radius (assuming a nearly isothermal profile at large radii). 
The two models shown in Fig.~\ref{Fig_BinnedProfile} (dashed lines) are truncated at a relatively large radius 
($10\deg$ or about 3.7\,Mpc at the distance of the cluster). 
The effect of the truncation radius can be noticed beyond a distance of about $8\deg$ from the centre, 
where both models start to converge towards zero.

The shallowness of the profile of Virgo in the outermost regions ($r > R_{\rm vir}$) 
may be due to the presence of significant amounts 
of plasma beyond the virial radius. The increased sensitivity of 
our analysis (due to the large area being integrated) allows us to explore the outskirts of a cluster to distances never examined before. 
For instance, the bin at a distance of about $4\deg$ from the centre integrates the signal over a total 
area of roughly $15$\,deg$^2$, and the bin at about $8\deg$ integrates a remarkable total area of 
$30$\,deg$^2$, reducing the noise levels to an unprecedentedly low value. 
The profile at large radii may be sensitive to the presence of the elusive WHIM, which is expected 
to contribute at the 1\,$\mu$K level around Virgo \citep{Fujimoto2004,Diego2008}. 
A reassuring result from Fig.~\ref{Fig_BinnedProfile} is the excellent agreement in the outer 
region ($r \approx 2R_{\rm vir}$) of the binned profile with the profile from the simulated cluster shown in Fig.~\ref{Fig_Sim_Virgo} (solid line in Fig.~\ref{Fig_BinnedProfile}). At smaller radii, the simulation systematically overpredicts the observed signal, which may be a consequence of insufficient modelling of the ongoing physical processes in the central region of the cluster (in particular a powerful AGN 
like M87 that may not be properly included in this simulation although the simulation does incorporate feedback from AGN activity).

\subsection{Contour-region bins}\label{ProfileSectors}
%########################
%__________________________________________________________________
\begin{figure} % MADE BY: CompareSignal_SZvsXray_Virgo_SNR_Contours_DX11.pro
   \centering
   \includegraphics[width=9.3cm]{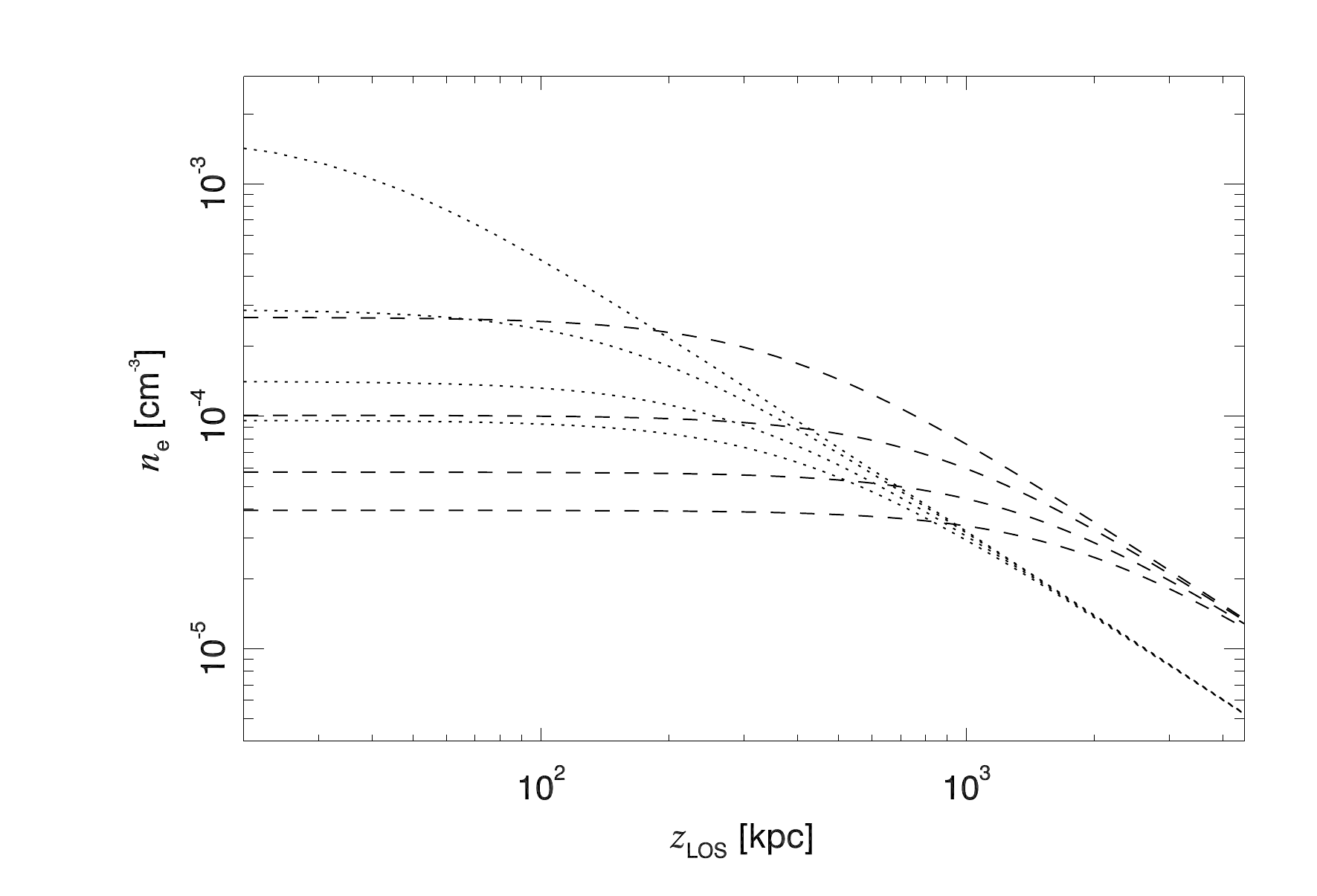}
      \caption{Model of the electron density along the line of sight coordinate (z$_{\rm LOS}$) for the 
               four different regions and for two sets of models. The dashed lines 
               are for a spherical model, while the dotted lines are for an ellipsoidal model. 
               For each group and from bottom to top, we show the cases for regions 1, 2, 3,  and 4.}
         \label{Fig_Regions_LOS}
   \end{figure}
%__________________________________________________________________
 
%__________________________________________________________________
\begin{figure} % MADE BY: CompareSignal_SZvsXray_Virgo_SNR_Contours.pro
   \centering
   \includegraphics[width=9.3cm]{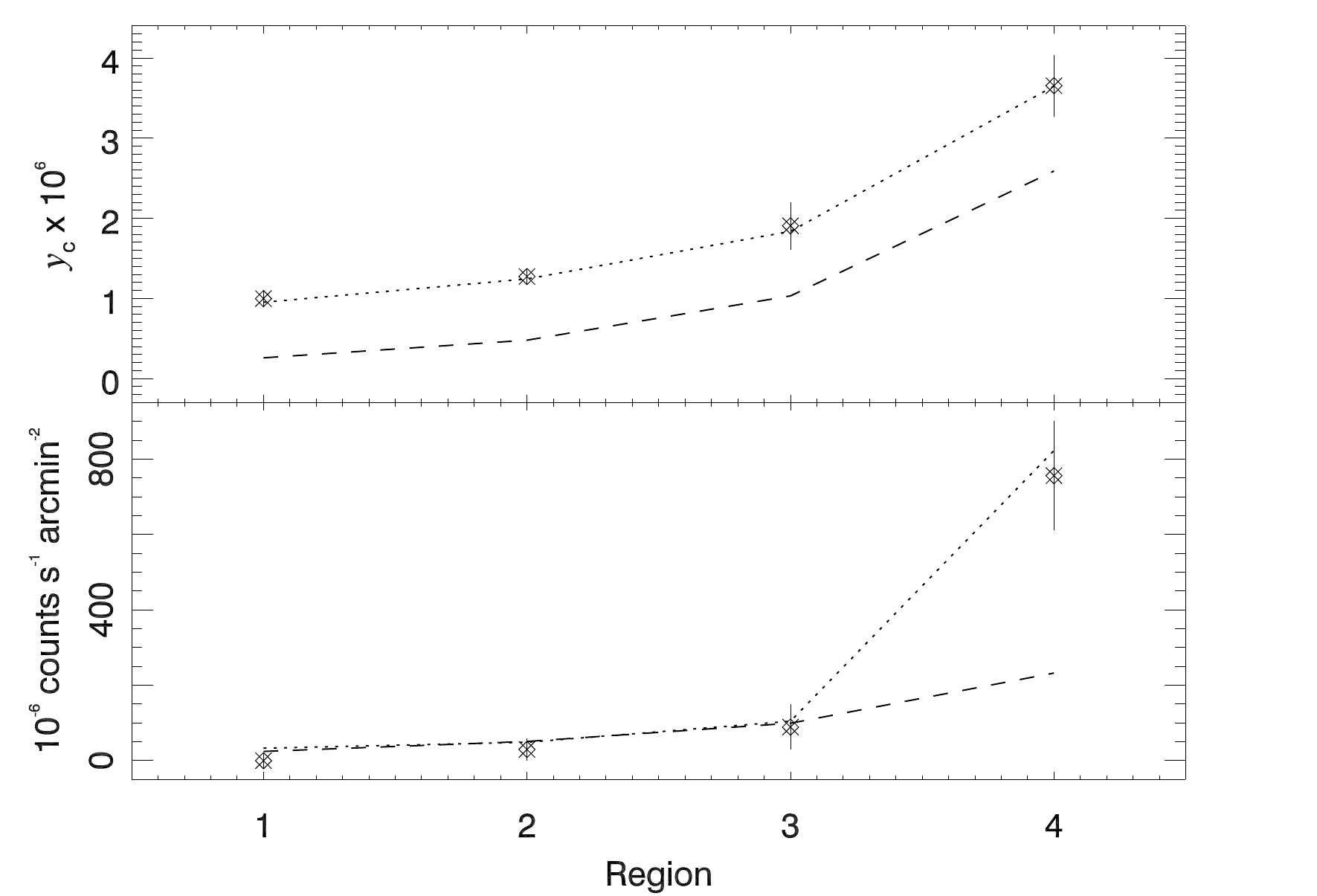}
      \caption{Mean signal (from the 15\arcm\ resolution maps) in each of the four regions 
               shown in Fig.~\ref{Fig_Regions} 
               (asterisks with error bars) compared with the predicted signal from analytical
                models. The models are shown in Fig.~\ref{Fig_Regions_LOS}, 
                as a dashed line for the spherical model and dotted line for the ellipsoidal model. 
                The error bars are the dispersion 
                in each region of the $1\pdeg5$ smoothed maps.}
         \label{Fig_Regions_data_model}
   \end{figure}
%__________________________________________________________________

The previous section presented results about the SZ cluster up to twice the virial radius. 
However, we made the assumption that the cluster is spherical when comparing this signal 
with the models, while clearly Virgo exhibits a non-spherical morphology. 
In order to test the possibility of constraining the plasma properties 
in the outskirts of the cluster independently of the geometry of the cluster,  
we divide the cluster into different regions according to the S/N in the SZ map shown in  
Fig.~\ref{Fig_Regions}. These regions are defined in terms of the dispersion of the smoothed SZ map. Region 
1 corresponds to the area where the signal in the smoothed map is between 2 and 3 times the dispersion of the smoothed 
map ($\sigma_{\rm s}$). Region 2 is for the signals between 3 and 4\,$\sigma_{\rm s}$, 
region 3 for signals  between 4 and 7\,$\sigma_{\rm s}$, and region 4 for signals larger than 7\,$\sigma_{\rm s}$. 
We retain only the regions that are associated with the cluster and, as in the 
previous section, we exclude the clump in the northern part of the Virgo cluster that is due to a 
Galactic residual (as discussed in Sect.~\ref{sect_GalactContam}). 
The regions shown in Fig.~\ref{Fig_Regions}, especially the outer ones, can be seen 
as areas of the cluster with similar optical depth (assuming the temperature does not 
change much, which should be a better assumption along lines of sight that do not cross the centre of the cluster). 
This is a consequence of the linear dependence of the SZ effect on the electron density and the relatively 
small change in the temperature. 
Although the regions are defined after smoothing the SZ map with a $1\pdeg5$ Gaussian kernel, 
we use the SZ map at the original 15\arcm\ resolution. 
To model each region we adopt a simple approach based on the  
observational X-ray constraints discussed above. The temperature is taken 
directly from the hybrid model discussed above. For the electron density, we adopt the $r^{-1.2}$ profile of U11, 
but assume that each region corresponds to the signal along a line 
of sight at an angular distance, or impact parameter $b$, from the centre. 
The normalization, $A$, is forced to take the same value for all the regions (to guarantee continuity 
in the electron density), while the impact parameter, $b$, is adjusted 
for each region in order to simultaneously reproduce the observed mean SZ and X-ray 
signal in each region, where the observed mean signal is defined as the total signal 
in a region divided by the number of pixels.

\begin{table*}[tmb]                 % table* is a two-column table.  Drop the * for one column.
\label{table1}
\begingroup
\newdimen\tblskip \tblskip=5pt
\caption{Parameters of the electron density profile in Eq.~\eqref{eq_ne_sector} for the different regions and 
         for two types of model. The first four rows are for the spherical model and the last four rows are for 
         the ellipsoid model. The parameter $b$ in the spherical model corresponds to the average 
         radius of a given region. The amplitude $A$ for this model is taken from the U11 model. 
         The parameters $A$ and $b$ in the ellipsoid model have been adjusted to 
         simultaneously fit the SZ and X-ray data. The relation between the two impact parameters 
         is well reproduced by the simple law $b_{\rm ellipsoid} = b_{\rm sphere}/4 - 50$. 
         The gas masses are computed for two different truncation radii, 2.5\,Mpc and 5\,Mpc.}
\label{tbl_1}    
\nointerlineskip
\vskip -3mm
\footnotesize
\setbox\tablebox=\vbox{
   \newdimen\digitwidth 
   \setbox0=\hbox{\rm 0} 
   \digitwidth=\wd0 
   \catcode`*=\active 
   \def*{\kern\digitwidth}
   \newdimen\signwidth 
   \setbox0=\hbox{+} 
   \signwidth=\wd0 
   \catcode`!=\active 
   \def!{\kern\signwidth}
%
%{\openup 1pt                                  % increase space between rows
{                                 % increase space between columns
\halign{\hfil#\hfil\tabskip=2em&                       % Template goes here.
        \hfil#\hfil&
        \hfil#\hfil&
        \hfil#\hfil&
        \hfil#\hfil&
        \hfil#\hfil\tabskip=0pt\cr
\noalign{\doubleline}
&& $A$& $b$& $M_{\rm gas}{(r<5\,\mathrm{Mpc})}$& $M_{\rm gas}{(r<2.5\,\mathrm{Mpc})}$\cr
 Symmetry & Region & [$10^{-5}$\,cm$^{-3}$]& [kpc]& [$10^{13}$\,M$_{\odot}$]& [$10^{13}$\,M$_{\odot}$]\cr
\noalign{\vskip 3pt\hrule\vskip 5pt}
Spheroidal & 1 &  8.5 & 1841 &  2.9 &  1.5\cr
           & 2 &  8.5 & 1343 &  3.6 &  2.4\cr
           & 3 &  8.5 & *840 &  4.8 &  3.6\cr
           & 4 &  8.5 & *374 &  7.0 &  5.9\cr
\noalign{\vskip 3pt\hrule\vskip 5pt}
Ellipsoidal & 1 & 3.2 &  *400 &  2.7 &  2.2\cr
            & 2 & 3.2 &  *290 &  3.0 &  2.6\cr
            & 3 & 3.2 &  *160 &  3.8 &  3.4\cr
            & 4 & 3.2 &  **37 &  6.3 &  5.9\cr
\noalign{\vskip 3pt\hrule\vskip 3pt}}}  % ends setbox and halign
}                                   % ends tabskip
%}                                   % ends openup
%\endPlancktable                    % ends one-column \halign
\endPlancktablewide                 % ends two-column \halign
%\tablenote a Footnote a.\par
%\tablenote b Footnote b.\par
\endgroup
\end{table*}                        % table* is a two-column table.  Drop the * for one column.

We parametrize the electron density as a function of the impact parameter, $b$, as 
\begin{equation}
n_{\rm e} = \frac{A}{\left(b^2 + z_{\rm LOS}^2\right)^{1.2/2}}\,,
\label{eq_ne_sector}
\end{equation}
where the coordinate $z_{\rm LOS}$ is integrated along the line of sight. 
In the above expression both $b$ and $z_{\rm LOS}$ are given in Mpc when $A$ takes the values of Table~\ref{table1}.  
Figure~\ref{Fig_Regions_LOS} shows the electron 
density profile used in each region along the line of sight.
We consider two scenarios. In the first one (dashed lines in Fig.~\ref{Fig_Regions_LOS}) 
we consider the cluster to have a spherically symmetric structure and that the U11 $r^{-1.2}$ estimation 
of the density profile is valid for the entire cluster (with the same normalization as in U11). 
In that case, the effective impact parameter, $b$, in Eq.~(\ref{eq_ne_sector}) should be close to 
the average radius in a given region. We call this case the ``spherical-like'' model. In the second scenario, 
(dotted lines in Fig.~\ref{Fig_Regions_LOS}) 
we assume a more realistic model where the spherical symmetry is not required. We still impose the constraint that, 
along a particular line of sight, the density has to follow an $r^{-1.2}$ law, but we relax the conditions on both 
the normalization, $A$, and impact parameter, $b$ (although $A$ is still forced to take the same value for all sectors). 
Letting the impact parameter take values that differ significantly 
from the average radius in a given region is similar to allowing the gas distribution to contract or expand 
along the line of sight in relation to the spherical distribution. We refer to this case as 
the ``ellipsoid'' model.
Values of $b$ significantly larger than the average radius in a region would 
correspond to density profiles that are more elongated along the line of sight (oblate). 
On the other hand, values of $b$ smaller than the average radius would correspond to a profile that 
has contracted along the line of sight in relation to the spherical case (prolate). 
The parameters from Eq.~\eqref{eq_ne_sector} derived from our fit in the spherical-like and 
ellipsoid models are shown in Table~\ref{tbl_1}. 
The models corresponding to these parameters are shown in Fig.~\ref{Fig_Regions_LOS}. 
The last two columns of Table~\ref{tbl_1} contain the estimated mass of the gas in each region for 
two assumptions about the truncation radius, defined as the maximum radius out to which we integrate the 
SZ or X-ray signals. 
These masses are derived adopting the non-spherical distribution of Virgo as follows. Given the model for the gas density 
along the line of sight (Eq.~\ref{eq_ne_sector}), the projected 
mass per pixel is computed for a single pixel in a given region and this mass is later multiplied by the number 
of pixels contained in that region. 
It is interesting to compare the masses for the two models. When the truncation radius is 2.5\,Mpc, the gas mass 
per region is very similar in both models, with the ellipsoid model being slightly more massive. The situation 
is reversed if the profiles are truncated at 5\,Mpc with the spherical model having more mass in this case.
This is a direct consequence of the higher density of the spherical model at $r>1$\,Mpc.

The predicted mean SZ and X-ray signal in each region is shown in Fig.~\ref{Fig_Regions_data_model} (as lines) compared with the actual measurement (symbols with error bars). 
Both models generally agree well with the X-ray data in regions 1--3.   
Less attention should be paid to the fit in region 4 (containing the centre of the cluster), since 
we are approximating all lines of sight in a sector by a single impact parameter. This approximation 
is a particularly bad one when the line of sight lies close to the centre (impact parameter $b \approx 0$). 
The models in this region are not expected to 
reproduce the data accurately, especially for the X-rays (due to the strong dependency with $n_{\mathrm{e}}^2$), 
since both models assume an impact parameter that misses the central cusp. 
For the SZ effect, the spherical model (which reproduces well the circularly averaged profile, as shown in the previous 
section) fails to reproduce the observed signal in all regions. This highlights the 
lack of symmetry of Virgo and the importance of considering non-spherical models when trying to reproduce 
the observations. 
When allowing the normalization $A$ and the impact parameter $b$ to change in order to 
accommodate both, SZ and X-ray data, we find 
that models that have smaller impact parameters, i.e., that concentrate the gas near the sky plane of 
the cluster (where the sky plane is defined as the plane perpendicular to the line of sight at Virgo's distance),  
are able to reproduce both SZ and X-ray region profiles simultaneously. A prolate model, where the axis 
of symmetry (the longest axis in the ellipsoid) is in the cluster plane (perpendicular to the 
line of sight), would be consistent with this picture. Our results disfavour 
oblate models, where the longest axis is parallel to the line of sight, since this would correspond to a much larger impact 
parameter $b$. This result is consistent with simulations and observations that conclude that clusters prefer 
to adopt prolate distributions \citep{Cooray2000,Sereno2006,Despali2014}. 

The increase in SZ signal when going from a spherical model to a prolate model (with 
approximately  the same amount of gas in both models) can be understood if we realize that the prolate model 
concentrates more gas closer to the cluster centre, where the temperature is higher, 
hence increasing the pressure. Understanding the relatively small change in the X-rays between our two models 
(in Table~\ref{table1}) is a bit more difficult due to the dependence on $n_{\rm e}^2$. However, we can see in 
Fig.~\ref{Fig_Regions_LOS}, how a relative compensation in the density between the central and outer regions  
may be the reason for the small change between the spherical and ellipsoid models. While the spherical model 
has a higher density at large radii, at small distances from the cluster plane the ellipsoid model has more density 
than the spherical model. Even though the range of distances where the ellipsoid model is denser than the spherical 
model is smaller, the X-ray surface brightness gets boosted by the $n_{\rm e}^2$ dependence.  
Also, in the case of X-rays, the weaker dependence on the temperature makes $T$ less relevant than in the SZ case. 

In the outermost region 1, our model predicts slightly more X-rays than are actually observed both for 
the spherical and ellipsoid models. A decrease in the X-ray signal while maintaining the SZ effect can be obtained if the 
electron density is reduced by a factor $p<1$ and the temperature is increased by a similar amount ($p^{-1} > 1$). 
The average SZ effect will remain the same while the X-ray average signal will decrease by a factor $p^{3/2}$. 
The small mismatch in region 1 could be an indication that the actual temperature in the outskirts of Virgo is 
higher than the one assumed in our model (1\,keV for $r> 1$\,Mpc).  
%The fact that the SZ agrees so well with a profile derived from $X$-ray data rules out a significant amount of clumpiness 
%in the plasma. Should clumpiness in Virgo be significant, it would bias the $X$-ray derived gas density towards 
%higher values that would result in a mismatch between the observed and predicted SZ ring-profiles. 
%The fact that clumpiness seem to be small (or negligible) in a relatively cold cluster like Virgo could 
%be extrapolated to other more massive clusters, where higher temperatures would make even harder the existence 
%of clumpiness in the plasma. 
Another mechanism that can increase the X-ray to SZ ratio is gas clumping. 
If the gas clusters in smaller clumps with enhanced local overdensities, this would result in an 
increase in the average X-ray emissivity while maintaining the averege SZ signal. While clumpiness can not be 
ruled out in the outer parts of the cluster (and may offer a possible explanation for the observed signal), 
the fact that smooth models can simultaneously reproduce the central region in both SZ and X-ray data, suggests 
that the clumpinees mechanism, if present, must be small in the central regions. 

Adopting the gas masses derived from the ellipsoid model in Table~\ref{table1}, the total gas mass of the plasma is 
$(1.4$--$1.6)\times 10^{14}$\,M$_{\odot}$. 
If we assume the cosmic value for the baryon fraction, $f_{\rm b}=\Omega_{\rm b}/\Omega_{\rm m} = 0.1834$ \citep{planck2014-a15},  
the total mass of the cluster would be  $(7.6$--$8.7)\times 10^{14}$\,M$_{\odot}$, consistent with the upper values 
estimated from galaxy dynamics, like those based on the flow of galaxies falling into Virgo, from which 
\cite{Karachentsev2014} recently estimated a total mass for Virgo of $(8 \pm 2.3)\times 10^{14}$\,M$_{\odot}$.

\subsection{Distance to the Virgo Cluster from SZ and $X$-ray data}\label{subsect_dist}
%-------------------------------------------------------------------------------------
In the above sections we have used the higher-end estimate of the distance to the Virgo cluster, 
motivated by the fact that we want to directly use recent estimates of the electron density in the central part 
of Virgo (as in G04). 
As noted in the introduction, the distance to the Virgo cluster is still a question of debate, with different 
values in the literature ranging from 16 to 21\,Mpc. Hence, it is interesting to derive a new distance estimation 
based on the combined SZ and X-ray data \citep{Silk1978,Cavaliere1979,Birkinshaw1991,Herbig1995}.

As described in Sect.~\ref{Sect_deproj}, one can derive the electron density and temperature from X-rays,  
assuming a given distance, and compare the X-ray-derived pressure with the observed SZ effect. 
Since the deprojected electron density is proportional to the inverse of the square root of the distance, 
an agreement between the predicted and observed pressure should be obtained when the correct distance is assumed. 
In  Fig.~\ref{Fig_PressureProfile} above, we show an example of how the inferred gas pressure 
depends on the distance. 
In this case, the difference between the two pressures for the two distances is relatively 
small. This is a direct consequence of the weak dependency of the inferred electron density with the 
square root of the distance, $\sqrt{D}$.

\begin{figure} % MADE BY: Plot_Distance_vs_GasMass.pro
   \centering
   \includegraphics[width=9.3cm]{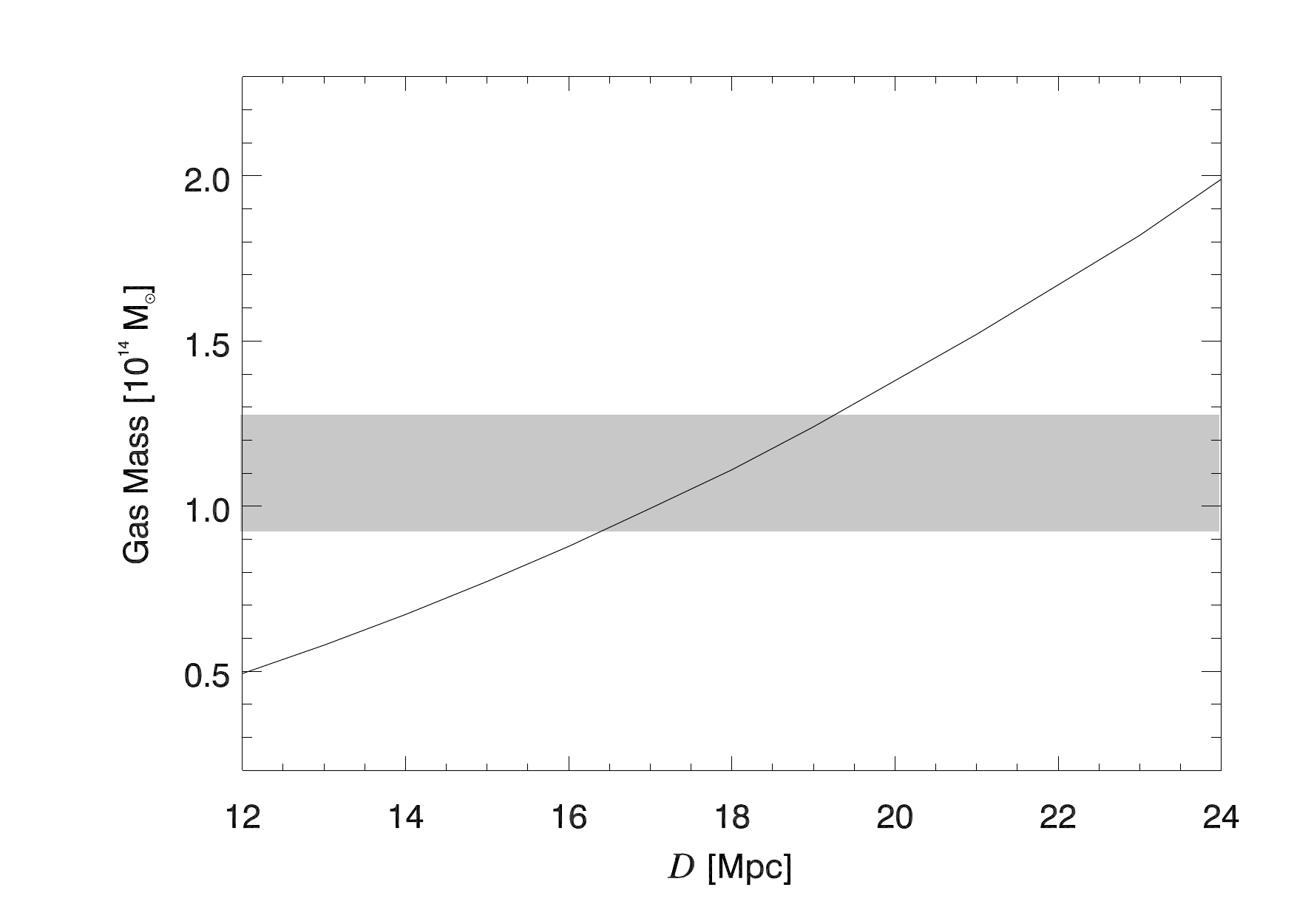}
      \caption{Estimated gas mass as a function of distance (solid line). 
               The shaded region corresponds to the gas mass derived assuming the total mass 
               of the cluster can take values from $5\times10^{14}$\,M$_{\odot}$ to $7\times10^{14}$\,M$_{\odot}$ 
               and that the fraction of mass corresponding to the gas 
               equals the cosmic value, $f_{\rm b}=\Omega_{\rm b}/\Omega_{\rm m} = 0.1834$. The shaded region assumes also 
               that different estimates of the total mass are independent of the distance.}
         \label{Fig_Mgas}
   \end{figure}

A more sensitive distance discriminator can be derived from the gas mass. 
The estimation of the gas mass depends on the volume element, which is different for different distances.  
This is better understood if we define the volume element as the volume subtended by a fixed solid angle, like a 
cubic pixel. Hence, while the SZ and X-ray predicted signals are basically the same for different distances 
(once the volume elements and conversion rates are re-scaled according to the new distance), the estimated 
gas mass changes with distance as $D^2$.   
Figure~\ref{Fig_Mgas} shows the dependence of the estimated mass of the gas on the distance to the cluster. 
As expected, the gas mass grows as the square of the 
distance (from the combination of the volume element, $\propto D^3$, and the integral along the line of 
sight of the individual volume element, $\propto D^{-1}$). When compared with the gas mass estimated 
from the total mass and the cosmic baryon fraction we can obtain a direct estimate of the distance to the cluster. 
Adopting a mass range for Virgo from $5\times10^{14}$\,M$_{\odot}$ to $7\times10^{14}$\,M$_{\odot}$, which is consistent 
with the range of masses discussed in the introduction
\citep{Vaucouleurs1960,Hoffman1980,Bohringer1994,Karachentsev2010,Karachentsev2014}, 
and the cosmic value $f_{\rm b}=\Omega_{\rm b}/\Omega_{\rm m} = 0.1834$, we can infer the gas mass 
(shaded band in Fig.~\ref{Fig_Mgas}) and compare it with the solid curve in the same figure.

%__________________________________________________________________
%__________________________________________________________________

This simple test suggests that the distance to Virgo is around 18\,Mpc, which fits nicely with the average 
estimate from different methods. Of course, this estimate relies on the assumption that the gas mass 
is known (or the total mass is known and the gas budget of the cluster up to the largest radii matches 
the cosmic value). If the distance (or redhift) is fixed, then the problem can be reversed to derive gas masses. 
This way of estimating distances to clusters could be of particular interest 
in future experiments (eROSITA, COrE+) where the estimation of the redshift of thousands of distant clusters 
may be a challenging task. Interestingly, this sensitivity to the redshift is a consequence of the relative 
insensitivity of the total SZ signal to redshift combined with the strong dependence of the total X-ray 
flux on redshift. 

It is useful to put the inferred distance to Virgo of around $18$\,Mpc in the context of earlier estimates 
of the Hubble parameter based on the distance to Virgo. \cite{Pierce1994} found (based on Cepheids) a relatively 
small distance to Virgo of 14.9\,Mpc and a corresponding $H_0 \approx 87$\,km\,s$^{-1}$\,Mpc$^{-1}$. On the opposite extreme, 
the largest distance estimates to Virgo, $D \approx 21$\,Mpc, come from 21-cm observations 
\citep{Sandage1976,Federspiel1998}, which also give a low estimate for $H_0 \approx 55$\,km\,s$^{-1}$\,Mpc$^{-1}$. 
Rescaling the relatively small or large distances found by these authors to 18\,Mpc would change the derived 
Hubble constant to $H_0 \approx 72$\,km\,s$^{-1}$\,Mpc$^{-1}$ (for \citealt{Pierce1994})  
or $H_0 \approx 64$\,km\,s$^{-1}$\,Mpc$^{-1}$ (for \citealt{Sandage1976,Federspiel1998}). 
These values are in better agreement with the value derived from the power spectrum of the CMB by \Planck\   
 (using TT, TE and EE constraints) of $H_0 = (67.6 \pm 0.6)$\,km\,s$^{-1}$\,Mpc$^{-1}$.  
It is also important to highlight that the earlier estimates of $H_0$ based on SZ+X-ray measurement 
found a value of $H_0 = (66 \pm 11)$\,km\,s$^{-1}$\,Mpc$^{-1}$ \citep{Jones2005}, also in good agreement 
with the latest estimate from \Planck.

%% _______________________________ SECTION CONCLUSION ____________________________
%_________________________________________________________________________________

%%%%%%%%%%%%%%%%%%%%%%%%%%%%%%%%%%%%%%%%%%%%%%
\section{Conclusions.}\label{sect_conclusion}
%%%%%%%%%%%%%%%%%%%%%%%%%%%%%%%%%%%%%%%%%%%%%%
\Planck's  full sky coverage, sensitivity, and wide frequency range offer the best opportunity to study the 
largest single SZ source in the sky, the Virgo cluster. 
The combination of SZ with X-ray observations 
is a powerful way of constraining the gas properties in galaxy clusters. 
The strong dependence of X-rays on the electron density is particularly useful to constrain the gas density. 
However, in order to constrain the gas properties several assumptions have to be made. 
One common assumption is that the gas distribution is smooth, that is, there is no 
significant clumpiness. When comparing the X-ray-derived profiles in rings from \cite{Ghizzardi2004} and \cite{Urban2011} 
with the \Planck\ SZ circular profile, we observe that the X-ray-derived profiles predict reasonably well 
the right amount of SZ signal without invoking the need for clumpiness to explain possible deviations. 
Another common assumption is that the cluster has spherical symmetry. 
We find that models derived from sparse \XMM\ data and a spherical deprojection are able 
to reproduce the ring-averaged signal relatively well, 
both in X-rays and in the SZ effect within the virial radius. 
However, when the irregular geometry of Virgo is taken into account, the 
spherical model predicts significantly less SZ signal than observed beyond the virial radius. 
A model that breaks the spherical symmetry and concentrates the gas closer to the cluster plane 
(where the temperature should be higher) produces a better fit, suggesting an ellipsoidal distribution 
for Virgo. A prolate-like distribution with the longest axis (or axis of symmetry) in the image plane, 
(i.e., perpendicular to the line of sight) would be consistent with our model. 
 
We study the signal in radial bins and find an excess in our SZ map reaching out to twice the 
virial radius. The observed signal is consistent with expectations from models based on an extrapolation 
of the electron density beyond the virial radius, with a power law $n_{\rm e} \propto r^{-1.2}$. 
The observed signal is also consistent with the expectation derived from a constrained simulation of 
Virgo. We observe that in order to reproduce the observed signal between $R_{\rm vir}$ and $2R_{\rm vir}$ with 
this power law, the temperature needs to be in the range of $1$\,keV beyond 1\,Mpc from the cluster centre. 
The properties of this rarefied medium coincide with those expected for the hottest phase of the WHIM.  
Finally, using the estimated total gas mass inferred from the combination of SZ and X-ray data and under 
the assumption that the total baryon fraction of Virgo is representative of the cosmic value, we infer 
a distance to Virgo of approximately $18$\,Mpc. 

\begin{acknowledgements}

%\section{Acknowledgments}
The Planck Collaboration acknowledges the support of: ESA; CNES, and
CNRS/INSU-IN2P3-INP (France); ASI, CNR, and INAF (Italy); NASA and DoE
(USA); STFC and UKSA (UK); CSIC, MINECO, JA and RES (Spain); Tekes, AoF,
and CSC (Finland); DLR and MPG (Germany); CSA (Canada); DTU Space
(Denmark); SER/SSO (Switzerland); RCN (Norway); SFI (Ireland);
FCT/MCTES (Portugal); ERC and PRACE (EU). A description of the Planck
Collaboration and a list of its members, indicating which technical
or scientific activities they have been involved in, can be found at
\url{http://www.cosmos.esa.int/web/planck/planck-collaboration}. Some of the results presented in this work are based on observations obtained 
with XMM-Newton, an ESA science mission with instruments and contributions directly funded by ESA Member States and NASA.
%\footnote{In particular we have used observations with the following ObsID; 
%0114120101, 0200920101, 0551870101, 0551870301, 0551870401, 0551870501, 0551870601, 0551870701, 0551871201, 0551871301, 0603260201, 
%0603260401, 0603260501, 0603260601, 0021540201, 0106060201, 0106060301, 0106060401, 0106060501, 0106060601, 0106060701, 0106060801, 
%0106060901, 0106061101, 0106061201, 0106061401, 0106860201, 0108260201, 0108860101, 0110930301, 0110930701, 0112550501, 0112550701, 
%0112550801, 0112551001, 0112552101, 0112610101, 0112840101, 0141570101, 0145800101, 0147610101, 0200130101, 0200650101, 0202730301, 
%0205010201, 0208020101, 0210270101, 0210270201, 0306060101, 0306630101, 0306630201, 0404120101, 0414980101, 0502050101, 0502050201, 
%0502160101, 0504100601, 0504240101, 0510011501, 0550540101, 0550540201, 0556210601, 0651790101, 0651790201, 0651790301, 0673310101, 
%0675140101} 

\end{acknowledgements}

\bibliographystyle{aat} % style aat.bst 
\bibliography{biblio,Planck_bib} % your references Yourfile.bib 

%%%%%%%%%%%%%%%%%%%%%%%%%%%%%%%%%%%%%%%%%%%%%%%%%%%%%%%%%%%%%%%%%%%%%%%%%%%%%%%%%%%%%%%
%%%%%%%%%%%%%%%%%%%%%%%%%%%%%%%%%%%%%%%%%%%%%%%%%%%%%%%%%%%%%%%%%%%%%%%%%%%%%%%%%%%%%%%
%%%%%%%%%%%%%%%%%%%%%%%%%%%%%%%%%%%%%%%%%%%%%%%%%%%%%%%%%%%%%%%%%%%%%%%%%%%%%%%%%%%%%%%

%%%%%%%%%%%%%%%%%%%%%%%%%%%%%%%%%%%%%%%%%%%%%%%%%%%
\begin{appendix}
\section{Stripe removal}\label{sect_stripes}
%%%%%%%%%%%%%%%%%%%%%%%%%%%%%%%%%%%%%%%%%%%%%%%%%%%
%==================================================

%__________________________________________________________________
\begin{figure*} % MADE BY: Destripe_SZILC_DX11.pro
   \centering
   \includegraphics[width=18.0cm]{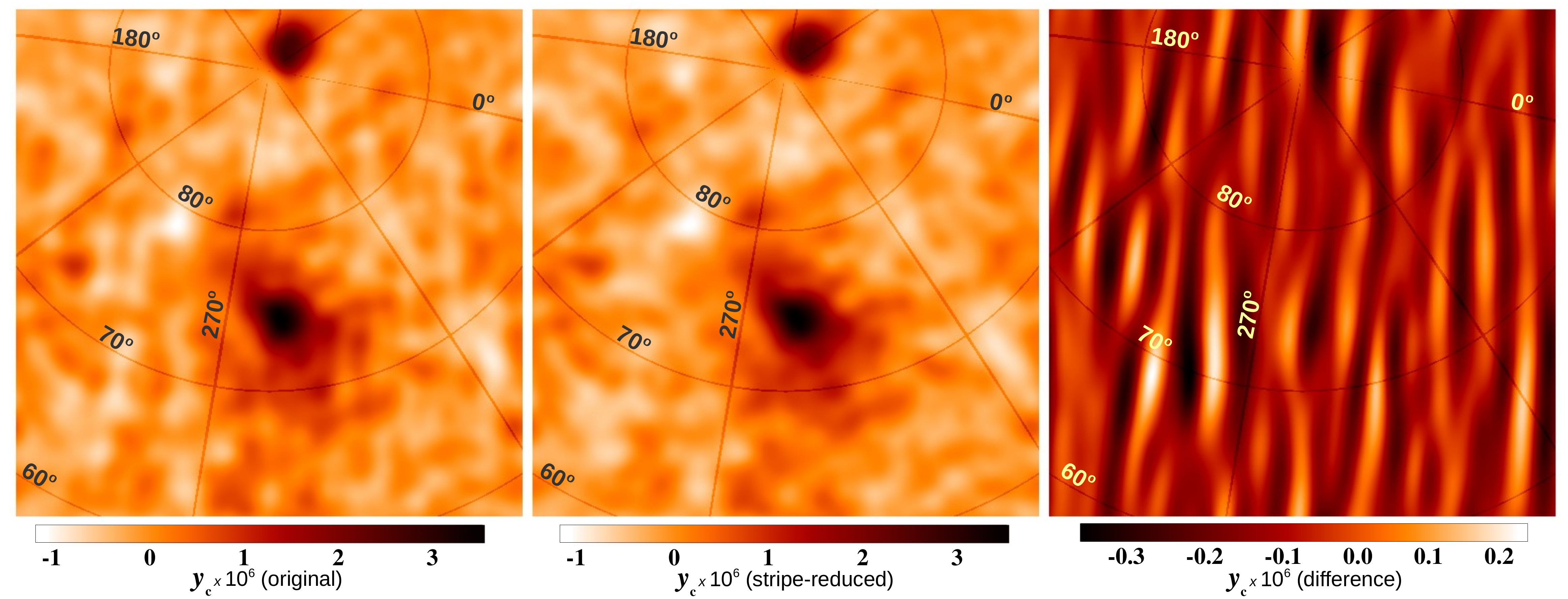}
      \caption{Estimation of the residual stripes in the SZ map in the Virgo region. All maps have been 
               degraded to a resolution of $1\pdeg5$. \emph{Left}: original data. \emph{Middle}: stripe-reduced data. 
               \emph{Right}: difference between the original and the stripe-reduced maps. 
               The colour bar in the right panel shows the difference in units of $10^7$ times the Compton parameter; the colour scale in the other panels is the same as in Fig.~\ref{Fig_SZ_Grid}.}
         \label{Fig_Stripes}
\end{figure*}
%__________________________________________________________________

One of the known artefacts of \Planck\ data (and present in other experiments) are residuals from 
baseline removal, or ``stripes''. These stripes can produce fluctuations of several microkelvin (at 150\,GHz) 
between neighboring scan trajectories. A discussion of the impact of stripes in the all-sky SZ map can be found also in \citep{planckSZallsky}. 
On small scales, these fluctuations can compete with and even dominate the weak SZ signals. 
Stripe residuals are small near the ecliptic poles where multiple scans overlap at different angles and they 
become larger near the ecliptic plane where the data are more noisy and the scan directions are nearly 
parallel to each other, making it harder to remove residual stripe signals. 
Since Virgo is located close to the ecliptic plane, stripes contaminate the SZ signal on 
small scales. These stripes are more evident in individual surveys, and hence individual survey maps can be used to isolate the 
stripe signal.
 
We produce SZ effect maps similar to the one described in the present paper, but based on individual surveys. 
The individual surveys are later combined in a destriping algorithm that removes stripe features. 
In the individual surveys, the stripe residuals tend to form elongated compact regions in Fourier space.
These regions can be identified for each survey and depend on the scanning pattern of each survey. 
For a given survey, the ``bad'' modes in Fourier space can be substituted by the corresponding 
good modes in a different survey that does not show this feature in the same region of Fourier space. 
Alternatively, the compact region containing the bad modes can be identifed in Fourier space and this region can 
be masked out.
 
When transforming the Fourier modes back into real space we recover a destriped version of that survey. 
The difference between the two maps (original and destriped version) is shown in Fig.~\ref{Fig_Stripes}. 
When smoothed with a 1\pdeg5 FWHM Gaussian, the stripes contribute only a small fraction of a 
microkelvin and the largest fluctuations do not affect the central region of Virgo. 
The effect of stripes is generally small on large scales and they tend to compensate each 
other when taking an average (as when computing profiles or smoothing the data). For instance, 
at 10\arcm\ resolution the stripes fluctuate, with amplitudes of about 4\,$\mu$K at 143\,GHz.
When the same SZ map is smoothed with a $1\pdeg5$ FWHM Gaussian, the amplitude of the stripes falls below 1\,$\mu$K.

\end{appendix}

\end{document}